\documentclass{aa}
\usepackage{natbib}
\usepackage{ucs}
\usepackage[utf8x]{inputenc}
\usepackage{amsfonts}
\usepackage{graphicx}
\usepackage{amssymb}
\usepackage{float}
\usepackage{amsmath}
\usepackage{hyperref}
\usepackage{fontenc}
\date{Accepted}
\begin{document}
\title{Investigating the cores of fossil systems with Chandra}
\author{V.~Bharadwaj\inst{1}, T.~H.~Reiprich\inst{1}, J.~S.~Sanders\inst{2}, G.~Schellenberger\inst{1}}
\institute{Argelander-Institut f\"ur Astronomie, Auf dem H\"ugel 71, 53121
  Bonn, Germany \and Max-Planck-Institut f\"ur Extraterrestriche Physik, Giessenbachstrasse 1, 85748, Garching, Germany}
\titlerunning{cores of fossil groups and clusters}
\abstract{}{We aim to systematically investigate the cores of a sample of fossil galaxy groups and clusters (``fossil systems'') using Chandra data to see what hints they can offer about the properties of the intracluster medium in these particular objects.}{We chose a sample of 17 fossil systems from literature with archival Chandra data and determined the cool-core fraction for fossils via three observable diagnostics, namely the central cooling time, cuspiness, and concentration parameter. We quantified the dynamical state of the fossils by the X-ray peak/brightest cluster galaxy (BCG) separation, and the X-ray peak/emission weighted centre separation. We also investigated the X-ray emission coincident with the brightest cluster galaxy (BCG) to detect the presence of potential thermal coronae. A deprojection analysis was performed for fossils with $z < 0.05$ to resolve subtle temperature structures, and to obtain the cooling time and entropy profiles. We also investigated the $L_{\mathrm{X}}-T$ relation for fossils from the 400d catalogue to test if the scaling relation deviates from what is typically observed for other groups.}{Most fossils are identified as cool-core objects via at least two cool-core diagnostics with the population of weak cool-core fossils being the highest. All fossils have their dominant elliptical galaxy within 50 kpc of the X-ray peak, and most also have the emission weighted centre within that distance. We do not see clear indications of a X-ray corona associated with the BCG unlike that has been observed for some other clusters. Fossils lack universal temperature profiles, with some low-temperature objects generally not showing features that are expected for ostensibly relaxed objects with a cool-core. The entropy profiles of the $z < 0.05$ fossil systems can be described well by a power law with shallower indices than what is expected for pure gravitational processes. Finally, the fossils $L_{\mathrm{X}}-T$ relation shows indications of an elevated normalisation with respect to other groups, which seems to persist even after factoring in selection effects.}{We interpret these results within the context of the formation and evolution of fossils, and speculate that non-gravitational heating, AGN feedback in particular, could have had an impact on the ICM properties of these systems.}
\authorrunning{V.~Bharadwaj et al.}
\maketitle
\section{Introduction}
Clusters of galaxies are the largest gravitationally bound structures in the Universe, with an aggregate mass between a few times $10^{13}~\mathrm{M_{\odot}}$ and $10^{15}~\mathrm{M_{\odot}}$. Consisting of galaxies, hot X-ray emitting gas (the intracluster medium-ICM) and dominated by dark matter, these objects are unique in their application to study both cosmology and astrophysics. On the lower mass end, these systems are sometimes called as groups, to indicate a smaller aggregation of galaxies. The distinction between groups and clusters is quite loose, with definitions in literature based on optical richness or simply a mass/temperature cut (e.g.~\citealt{2012MNRAS.422.2213S}). 

Within the existing division of groups and clusters, lie an ostensibly `special' class of systems called as fossil systems; systems which are dominated by a single, bright, elliptical galaxy and deficient in other bright galaxies. The first fossil group was reported by \cite{1994Natur.369..462P} who showed the existence of a system at redshift 0.171 which had a large, X-ray halo associated with an elliptical galaxy. The choice of nomenclature `fossil' is used to indicate a class of systems which is a remnant of galaxy merging, with hot X-ray gas and an elliptical galaxy the only remnants of this process. A `formal' definition for a fossil system was provided by \cite{2003MNRAS.343..627J} who defined it as a spatially extended X-ray source with luminosity larger than $10^{42} h^{-2}_{50}$ erg/s with an optical counterpart wherein the magnitude difference between the first and second brightest galaxy is greater than or equal to 2 in the Johnson R band. Simulations by \cite{2007MNRAS.382..433D} show that systems selected by this criterion represent a class of objects which assemble a higher fraction of their mass at high redshift as compared to non-fossils, further justifying their tag of being early-forming `old' systems. This formal definition, however, is open to contention as pointed out again by \cite{2010MNRAS.405.1873D} who argue via simulations that a magnitude difference of 2.5 between the first and fourth brightest galaxy identifies 50\% more early formed systems. 

The most widely acknowledged formation scenario for these systems is one wherein dynamical friction causes galaxies close to the centre of the group to merge and leave behind a large elliptical galaxy and a hot X-ray halo (e.g.~\citealt{2005ApJ...630L.109D}). An outstanding question is if the properties of fossil systems make them a \textit{special} class of systems. Also, it is still unclear whether they are the final stages of mass assembly or are merely a temporary state in the formation of larger clusters. \cite{2009AJ....137.3942L} argue that fossils are not special, and merely represent the final stages of mass assembly in a region without enough surrounding matter. Simulations by \cite{2008MNRAS.386.2345V} claim however, that the `fossil' phase is a temporary one and the magnitude gap would end by renewed infall from the surroundings. Adding to the confusion are results by \cite{2011MNRAS.418.2054P} who argue that the high mass to light ratios of these systems and their low richness are not predicted by any current model of fossil formation. \cite{2011MNRAS.417.2927P} argue that compact groups are the progenitors of low-mass fossils, while more massive fossil systems are formed by an early infall of massive satellites, hinting that a singular explanation for the formation of fossils might be inadequate.

Very few fossil systems have been thoroughly studied, which adds to the lack of sound conclusions on their formation and evolution (e.g.~\citealt{1999ApJ...514..133M, 2004MNRAS.349.1240K,2004ApJ...612..805S}). X-ray sample studies of fossil systems, which could provide us additional details, are even rarer (e.g.~\citealt{2007MNRAS.377..595K,2012ApJ...747...94M,2012ApJ...752...12H}) and have focused more on global properties such as X-ray luminosity, temperature and mass. Till date, there has never been a detailed and focused investigation of the cores of a sample of fossil systems in X-rays, particularly with respect to their cool-core properties.

The cores of galaxy clusters and groups offer important hints on the dynamical state of the ICM in the central regions, e.g. systems which are morphologically disturbed, generally have long central cooling times (CCTs) and elevated central entropies as compared to relaxed systems which show much shorter CCTs and lower central entropies. The clusters with the strongest cooling, namely the strong cool core (SCC) clusters, show very short CCTs ($\leq$ 1 Gyr), low entropies ($\leq$ 30 keV $\mathrm{cm}^{2}$), and central temperature drops \citep{2010A&A...513A..37H}. This is not necessarily true on the group regime, where there are examples of systems which do not show any indication of cool gas in the cores despite short CCTs and low entropies (e.g.~\citealt{2014A&A...572A..46B}). Feedback also plays an important role in cluster/group cores as without it, there would be catastrophic cooling with very high mass deposition rates, inconsistent with observations (e.g.~\citealt{2001A&A...365L.104P,2001A&A...365L..87T,2001ASPC..234..351K,2003ApJ...590..207P,2008MNRAS.385.1186S}). Where do fossil systems fit in this picture? What hints can the cool core properties of these systems offer in terms of their formation and evolution? These are the questions we attempt to address in this paper.

This paper is organised as follows: Section~\ref{Data} deals with the data and analysis. We present our results and discuss them in section~\ref{ResD}. A short summary is presented in section~\ref{Summary}. Throughout this work, we assume a $ \Lambda$CDM cosmology with $ \Omega_{\mathrm{m}} = 0.30$, $\Omega_{\Lambda} = 0.70$ and $ h = 0.70$ where $H_{0} = 100 h$ km/s/Mpc, unless stated otherwise. All errors are quoted at the $68\%$ level. Log is always base 10 here.

\section{Data and Analysis}\label{Data}
\subsection{Sample}
For this study, we selected a sample of 17 fossil systems and candidates spanning a large redshift range (from $\sim 0.01$ to $\sim 0.4$) with good quality archival Chandra data ($t_{\mathrm{exp}} \geq 10$ ks) as required for exploring the properties under consideration here. Chandra was preferred over XMM-Newton due to its superior spatial resolution which makes it easier to probe the central regions which is the major focus of the work. Though the systems were categorized as fossil systems through different techniques, qualitatively the basic selection criteria of a lower X-ray luminosity threshold and a magnitude difference between the first and second brightest galaxies, is maintained for all of them. We also endeavoured to study the $L_{\mathrm{X}}-T$ relation for those fossil systems selected from the 400d cluster sample (Sec.~\ref{LTsec}) for which we analysed XMM-Newton data for one more fossil group, which was not included for the other studies. Table~\ref{samptab} summarises the basic information of the sample with redshifts and literature sources.

\begin{table*}
\caption{The sample of fossil systems. The starred entry represents the system which was only considered in the $L_{\mathrm{X}}-T$ study. Columns are (1) name of the fossil system, (2) redshift (3) literature sources.}\label{samptab}
\centering
 \begin{tabular}{|c|c|c|}
  \hline
 Name & redshift & Literature\\ \hline \hline
  NGC~6482 & 0.0131 &\cite{2004MNRAS.349.1240K} \\
  NGC~1132 & 0.0232 &\cite{1999ApJ...514..133M}/\cite{2004AdSpR..34.2525Y} \\
  RX~J0454.8-1806 & 0.0314 &\cite{2004AdSpR..34.2525Y} \\
  ESO~306-G 017 & 0.0358 &\cite{2004ApJ...612..805S} \\
  UGC~842 & 0.045 &\cite{2008ApJ...684..204V}/\cite{2010AJ....139..216L} \\
  RX~J1331.5+1108 & 0.081 &\cite{2003MNRAS.343..627J} \\
  cl1159p5531 & 0.081 &\cite{1999ApJ...520L...1V}/\cite{2010ApJ...708.1376V} \\
  cl2220m5228 & 0.102 &\cite{2010ApJ...708.1376V} \\
  $\mathrm{cl1038p4146}^{*}$ & 0.125 &\cite{2010ApJ...708.1376V} \\
  cl1416p2315 & 0.138 &\cite{2003MNRAS.343..627J}/\cite{2010ApJ...708.1376V} \\
 cl0245p0936 & 0.147 &\cite{2010ApJ...708.1376V} \\
 cl1340p4017 & 0.171 &\cite{1994Natur.369..462P}/\cite{2010ApJ...708.1376V} \\
  RX~J2247.4+0337 & 0.199 &\cite{1999ApJ...520L...1V}/\cite{2010ApJ...708.1376V} \\
  RX~J0825.9+0415 & 0.225 &\cite{2009AN....330..978E} \\
  RX~J1256.0+2556 & 0.232 &\cite{2003MNRAS.343..627J} \\
  RX~J0801+3603 & 0.287 &\cite{2009AN....330..978E} \\
  RX~J1115.9+0130 & 0.352 &\cite{2009AN....330..978E} \\
  RX~J0159.8-0850 & 0.405 &\cite{2009AN....330..978E} \\ \hline
 
 \end{tabular}

\end{table*}

\subsection{Basic data reduction}
Most of the data reduction steps were similar to that followed in \cite{2015A&A...573A..75B} and \cite{2014A&A...572A..46B} and we describe them here briefly. CIAO 4.4\footnote{\url{http://cxc.harvard.edu/ciao}} with CALDB 4.5.0 was used for the data reduction for all but one object--cl1416p2315. While the manuscript was under review, a very recent (2014), and much deeper observation of cl1416p2315 was made available in the Chandra archives which we analysed with CIAO 4.7 and CALDB 4.6.7, to account for changes in the ACIS time-dependent gain and to use the proper blank-sky background files. This is unlikely to have an effect on other objects in our sample which have much older observations, and is already accounted for in CALDB 4.5.0, and therefore we decided not to re-analyse them. Other than these changes, all the steps that were carried out were exactly the same.

The \verb+chandra_repro+ task reprocessed the data and removed afterglows, created the bad-pixel table and applied the calibration. The \verb+lc_clean+ algorithm was used to filter the soft proton flares, with the lightcurves visually inspected to check for any residual flaring. Point sources were detected and excluded from further analysis with the \verb+wavdetect+ wavelet algorithm. Since our focus is on the bright, central regions, background subtraction for the spectral and surface brightness analysis was done throughout using the blank sky background files. The X-ray emission peak (EP) was determined in an exposure corrected, point source subtracted image using the CIAO tools built into ds9 with a pixel scale of 2.

For extracting spectra, we created annular regions with a minimum source counts threshold of 1000. This ensured that the annuli are sufficiently small and at the same time does not sacrifice too much on the signal to noise. The resultant spectra were fit with an absorbed APEC model in the 0.7--7.0 keV energy range, with the $n_{\mathrm{H}}$ value taken from the built in $n_{\mathrm{H}}$ ftools which outputs the value from the Leiden/Argentine/Bonn (LAB) survey \citep{2005A&A...440..775K}. The abundance was kept thawed for the spectral fit and in cases where we could not get a reasonable constraint on the value, we froze it to 0.30. We used the \cite{1989GeCoA..53..197A} abundance table throughout.

The surface brightness profile (SBP) was obtained by centring on the EP in the 0.5--2.0 keV energy band. This SBP was then fit by a single or double beta model (e.g.~\citealt{1976A&A....49..137C}) given by:
\begin{equation}
\Sigma = \Sigma_{0} \left [ 1+\left ( \frac{x}{x_{\mathrm{c}}} \right )^{2} \right ]^{-3\beta+1/2} \\
\end{equation}

\begin{equation}
\Sigma = \mathrm{\Sigma}_{01}\left[1+\left(\frac{x}{x_\mathrm{{c_{1}}}}\right)^{2}\right]^{-3\mathrm{\beta}_{1}+1/2} +  \mathrm{\Sigma}_{02}\left[1+\left(\frac{x}{x_\mathrm{{c_{2}}}}\right)^{2}\right]^{-3\mathrm{\beta}_{2}+1/2}  
\end{equation}
where $x_{\mathrm{c_{i}}}$ is the core radius. This in turn gave us the electron density profile for the single and double beta cases as:

\begin{eqnarray}
\centering
n &=& n_{0}\left [ 1+\left ( \frac{r}{r_{\mathrm{c}}} \right )^{2} \right ]^{\frac{-3\beta}{2}} \\
n &=& \left(
n_{01}^2\left[1+\left(\frac{r}{r_\mathrm{{c_{1}}}}\right)^2\right]^{-3\beta_1} +
n_{02}^2 \left[1+\left(\frac{r}{r_\mathrm{{c_{2}}}}\right)^2 \right]^{-3\beta_2}
\right)^{1/2}
\end{eqnarray}
where $r_{\mathrm{c}}$ is the physical core radius. The central electron density $n_{0}$ for the single beta case is given by:

\begin{equation}
\centering
 n_{0} = \left ( \frac{10^{14}4\pi D_{\mathrm{A}}D_{\mathrm{L}}\zeta N}{\mathrm{EI}} \right )^{\frac{1}{2}} 
\end{equation}
Here, $N$ is the normalisation of the APEC model in the innermost annulus, $\zeta$ is the ratio of electrons to protons ($\sim 1.2$), $D_{\mathrm{A}}$ is the angular diameter distance and $D_{\mathrm{L}}$ is the luminosity distance. $\mathrm{EI}$ is defined as:

\begin{equation}
 \mathrm{EI} = 2\pi  \int_{-\infty }^{\infty } \!  \int_{0}^{R} x\left(1+\frac{x^{2}+l^{2}}{x^{2}_{c}} \right)^{-3\beta} \mathrm{d}x\mathrm{d}l                                                                                                                                                                                                                                                                                                       
\end{equation}
where $R$ is the radius of the innermost annulus.

Similarly, for the double beta case, $n_{0}$ (Appendix A of \citealt{2010A&A...513A..37H}) is given by:

\begin{equation} 
n_{0} = \left [\frac{10^{14}4\pi (\Sigma_{12}\mathrm{LI_{2}}+\mathrm{LI_{1}} )D_{\mathrm{A}}D_{\mathrm{L}}\zeta N}{\Sigma_{12}\mathrm{LI}_{2}\mathrm{EI}_{1}+\mathrm{LI}_{1}\mathrm{EI}_{2} }  \right ]^{\frac{1}{2}}
\end{equation}

with the same definitions as before. $\mathrm{LI}_{i}$ is the line emission measure and is defined as:
\begin{equation}
 \mathrm{LI}_{i} = \int_{-\infty}^{\infty} \left(1+\frac{l^{2}}{x^{2}_{c_{i}}} \right)^{-3\beta_{i}} \mathrm{d}l
\end{equation}

To estimate the uncertainty on $n_{0}$, we conducted Monte Carlo simulations where the surface brightnesses were varied within their errors to generate new SBPs. These new SBPs were then fit again to a single or double beta model which was then used to determine $n_{0}$ (the APEC normalisation is also varied each time within the errors). The process was repeated 500 times to get a distribution for the values of $n_{0}$, and the standard deviation of the distribution gives the uncertainty on the measured $n_{\mathrm{0}}$.

\subsection{Cool-core analysis}
The most robust parameter to identify the cool-core nature of an object is the CCT \citep{2010A&A...513A..37H}, a quantity that is dependent on the central temperature and the central density. But, considering that most (59\%) of our objects have a redshift $\geq 0.1$, the determination of the CCT becomes problematic given that it is difficult to resolve the temperature profile for these objects in the central regions. Moreover, in some cases the data quality is not good enough to determine a temperature profile. Thus, in addition to the CCT, we used two other diagnostics, namely, Cuspiness $\alpha$ and concentration parameter $c_{\mathrm{SB}}$, both of which have been shown to have a strong correlation with the cooling time (e.g.~\citealt{2010A&A...521A..64S,2010A&A...513A..37H}). Indeed, recent studies of high redshift systems have used these two quantities to identify their CC nature (e.g.~\citealt{2012ApJ...761..183S,2015MNRAS.447.3723P}). In our study, if a fossil satisfied two out of three CC diagnostics, it was classified as a CC system.

The CCT is given as:
\begin{equation}
\label{CCTeq} 
\mathrm{CCT} = t_\mathrm{{cool}}(0) = \frac{3}{2}\zeta \frac{(n_\mathrm{{e0}}+n_\mathrm{{i0}})kT_{0}}{{n}^{2}_\mathrm{{e0}}\Lambda(T_{0},Z_{0})}  
\end{equation}
where ${n}_{i0}$ and ${n}_{e0}$ are the central ion and electron
densities, respectively, and $T_{0}$ is the temperature in the innermost annulus. As with \cite{2010A&A...513A..37H} and \cite{2014A&A...572A..46B} we took the value of central density values to be the value at $r=0.004r_{500}$. To estimate $r_{500}$, we first determined the virial temperature ($T_{\mathrm{vir}}$) in a single region centred on the EP, which extended to the outermost boundary of the outermost temperature bin. The size of this region effectively corresponds to 0.3-0.5$r_{500}$ for most objects, where the temperature gradient is not too significant (see Fig.~\ref{EPEWC_cusp} right, appendix \ref{Tempprofiles}). We then used the following scaling relation by \cite{1996ApJ...469..494E} to determine $r_{500}$ as was done in \cite{2010A&A...513A..37H} and \cite{2014A&A...572A..46B}:

\begin{equation}
r_{500} = 2 \times \left( \frac{T_{\mathrm{vir}}}{10~\mathrm{keV}} \right)^{\frac{1}{2}} \mathrm{Mpc} 
\end{equation}

To determine $T_{\mathrm{vir}}$ for Cl1038p4146 we used the XMM-Newton data sets which were analysed with the standard pipelines as illustrated in \cite{2015A&A...575A..30S}. Spectrum was once again extracted in a single region centred on the EP. Note that temperatures determined from Chandra and XMM-Newton can differ significantly as shown by \cite{2015A&A...575A..30S} and hence to remain consistent, we used the formula given in that study to convert the XMM-Newton temperature to Chandra temperature:
\begin{equation}
 \mathrm{log}_{10}(T_{\mathrm{XMM}}) = 0.889 \times \mathrm{log}_{10}(T_{\mathrm{Chandra}})
\end{equation}

There are five fossil systems for which we could not obtain a temperature profile. Hence, for these objects we assumed $T_{\mathrm{vir}}$ to be the central temperature while constraining the CCT. To estimate the uncertainties on the CCT for these objects, we conservatively assumed the lower bound of the temperature to be 0.6 times of $T_{\mathrm{vir}}$ (the largest central temperature decrement observed for nearby fossils, i.e. ES0306) and the upper bound to be 1.25 times $T_{\mathrm{vir}}$ (the largest central temperature increment, i.e. NGC 6482). A CC system has a CCT $< 7.7$ Gyr.

The Cuspiness $\alpha$ \citep{2007hvcg.conf...48V} is defined as:
\begin{equation}
 \alpha = -\frac{\mathrm{d~log}(n)}{\mathrm{d~log}(r)}
\end{equation}
which in model parameters \citep{2010A&A...513A..37H} for the single beta model is expressed as:
\begin{equation}
 \alpha = \frac{3\beta r^{2}}{r_{\mathrm{c}}^2 + r^{2}}
\end{equation}
and for the double beta model is:
\begin{equation}
 \alpha = 3r^{2} \frac{\Sigma_{12}\mathrm{LI}_{2}\beta_{1}r^{-2}_{\mathrm{c}_{1}}b^{'}_{1} + \mathrm{LI}_{1}\beta_{2}r^{-2}_{\mathrm{c}_{2}}b^{'}_{2}}{\Sigma_{12}\mathrm{LI}_{2}b_{1} + \mathrm{LI}_{1}b_{2}}
\end{equation}
Here, $\Sigma_{12}, \mathrm{LI}_{1}$ and $\mathrm{LI}_{2}$ are as defined for the CCT, $b_{i}$ and $b^{'}_{i}$ are defined as:
\begin{equation}
 b_{i} \equiv \left ( 1 + \left(\frac{r}{r_{c_{i}}}\right)^{2} \right )^{-3\beta_{i}}
\end{equation}
and 
\begin{equation}
 b^{'}_{i} \equiv  \left ( 1 + \left(\frac{r}{r_{c_{i}}}\right)^{2} \right )^{-3\beta_{i}-1}
\end{equation}
respectively. In both single and double beta cases, $r = 0.04r_{500}$. A CC system has $\alpha \geq~0.5$.
The error on $\alpha$ is once again determined by Monte Carlo simulations as with $n_{0}$.

The concentration parameter $c_{\mathrm{SB}}$ \citep{2008A&A...483...35S} is given as:
\begin{equation}
 c_{\mathrm{SB}} \equiv  \frac{\Sigma(r< \mathrm{40~kpc})}{\Sigma(r< \mathrm{400~kpc})}
\end{equation}
where $\Sigma$ is the integrated surface brightness within the defined radii (40 and 400 kpc). A CC system has $c_{\mathrm{SB}}~\geq~0.075$. As the $z < 0.05$ fossils are too close for estimating the surface brightness within 400 kpc, we did not determine $c_{\mathrm{SB}}$ for these systems.

\section{Results and Discussion}\label{ResD}
 \subsection{Cool core properties}
\begin{table*}
 \caption{Cool-core diagnostics of the fossils. The columns are (1) name of system, (2) coordinates (J2000) of the EP, (3) virial temperature of fossil, (4) central cooling time, (5) cuspiness, (6) concentration parameter, (7) cool-core or not, via any two diagnostics. Starred entries represent fossils for which the CCT was determined using $T_{\mathrm{vir}}$.}
  \renewcommand{\arraystretch}{1.2}
  \begin{tabular}{|c c c c c c c|}
  \hline \hline
   Name & EP (RA/DEC) &$T_{\mathrm{vir}}$ (keV) &CCT (in Gyr)&$\alpha$&$c_{\mathrm{SB}}$& CC or not?\\ \hline
 NGC~6482 &$17:51:48.85~+23:04:19.34$&$0.75^{+0.0078}_{-0.0079} $ &$0.124^{+0.015}_{-0.013} $ &$1.46^{+0.01}_{-0.01} $ &- &YES\\
 NGC~1132 &$02:52:51.79~-01:16:29.13$&$1.23^{+0.015}_{-0.016} $ &$1.18^{+0.18}_{-0.14}$ &$0.969^{+0.066}_{-0.066}$ &- & YES\\
  RX~J0454.8-1806 &$04:54:52.34~-18:06:55.57$&$2.36^{+0.07}_{-0.07} $ &$2.81^{+1.38}_{-0.78} $ &$0.888^{+0.020}_{-0.020} $ &- &YES \\
  ESO~306-G 017 &$05:40:06.75~-40:50:10.80 $&$2.71^{+0.07}_{-0.05} $ &$0.497^{+0.064}_{-0.057}$ &$0.790^{+0.022}_{-0.022} $ &- &YES \\
  UGC~842 &$01:18:53.67~-01:00:06.75$&$1.91^{+0.07}_{-0.07} $ &$1.75^{+0.22}_{-0.18}$ &$1.02^{+0.06}_{-0.06}$ &- &YES\\
  RX~J1331.5+1108 &$13:31:29.65~+11:07:57.44$&$0.74^{+0.02}_{-0.03} $ &$1.01^{+0.38}_{-0.23}$ &$1.33^{+0.11}_{-0.11}$ &$0.110^{+0.006}_{-0.006}$ &YES\\
  cl1159p5531 &$11:59:52.20~+55:32:06.29$&$1.70^{+0.06}_{-0.02} $ &$0.261^{+0.032}_{-0.028}$ &$1.78^{+0.20}_{-0.20}$ &$0.285^{+0.005}_{-0.005}$ &YES \\
  $\mathrm{cl2220m5228}^{*}$ &$22:20:08.64~-52:27:50.33$&$4.12^{+0.38}_{-0.35} $ &$20.9^{+2.74}_{-2.34} $ &$0.264^{+0.037}_{-0.037} $ &$0.062^{+0.006}_{-0.006}$ &NO\\
  cl1416p2315&$14:16:27.39~+23:15:22.59 $&$3.98^{+0.14}_{-0.14} $ &$4.73^{+1.54}_{-1.01}$ &$0.548^{+0.053}_{-0.053} $ &$0.059^{+0.002}_{-0.002} $ &YES\\
  $\mathrm{cl0245p0936}^{*}$&$02:45:48.83~+09:36:37.30 $&$2.45^{+0.38}_{-0.31}$ &$4.99^{+0.90}_{-0.76} $ &$0.790^{+0.028}_{-0.028}$&$0.116^{+0.012}_{-0.012}$ &YES \\
  $\mathrm{cl1340p4017}^{*}$ &$13:40:32.70~+40:17:39.26 $&$1.42^{+0.12}_{-0.10}$ &$3.29^{+0.53}_{-0.44} $ &$0.620^{+0.025}_{-0.025}$&$0.179^{+0.011}_{-0.011} $ &YES\\
  $\mathrm{RX~J2247.4+0337}^{*}$ &$22:47:27.88~+03:37:00.22$&$2.43^{+0.44}_{-0.35} $ &$15.9^{+3.58}_{-2.84} $ &$0.359^{+0.036}_{-0.036}$ &$0.060^{+0.006}_{-0.006} $ &NO\\
  RX~J0825.9+0415 &$08:25:57.83~+04:14:47.12$&$4.48^{+0.34}_{-0.29} $ &$5.16^{+0.76}_{-0.55}$ &$0.460^{+0.071}_{-0.071} $ &$0.086^{+0.006}_{-0.006}$ &YES\\
  $\mathrm{RX~J1256.0+2556}^{*}$ &$12:56:02.25~+25:56:35.38 $&$3.13^{+0.75}_{-0.51} $ &$24.2^{+5.15}_{-3.84}$ &$0.250^{+0.045}_{-0.045}$ &$0.074^{+0.010}_{-0.010} $ &NO\\
  RX~J0801+3603 &$08:00:56.81~+36:03:23.67$&$7.43^{+0.29}_{-0.30} $ &$2.25^{+0.32}_{-0.24}$ &$0.822^{+0.039}_{-0.039} $ &$0.120^{+0.003}_{-0.003}$ &YES\\
  RX~J1115.9+0130 &$11:15:51.80~+01:29:55.48$&$7.12^{+0.26}_{-0.26} $ &$0.629^{+0.039}_{-0.037} $ &$1.20^{+0.03}_{-0.03}$ &$0.224^{+0.005}_{-0.005}$ &YES\\
  RX~J0159.8-0850 &$01:59:49.36~-08:49:59.45$&$8.46^{+0.39}_{-0.33}$ &$0.598^{+0.040}_{-0.038}$ &$1.10^{+0.02}_{-0.02} $ &$0.223^{+0.006}_{-0.006} $ &YES\\ \hline
  
   \end{tabular}
\label{CCprop}

\end{table*}

Table~\ref{CCprop} summarises the CC properties of the fossil systems. We find that 14 out of 17 systems, i.e., 82\% are cool core systems, as evidenced by at least two out of the three CC diagnostics. If one uses the CCTs, we observe a range of values from a minimum of 0.124 Gyr to a maximum of 24.2 Gyr. The sub-classifications of the fossil systems using just the CCTs are 29\% SCC (CCT~$<~1$ Gyr), 53\% WCC ($1~\leq~$CCT~$<~7.7$ Gyr), and 18\% NCC (CCT $\geq 7.7$ Gyr).

As we have already pointed out, it is not always possible to determine the CCT and this becomes especially true for X-ray survey data (e.g. eROSITA). Thus, we also state the effect of using one CC parameter alone. With only the cuspiness, we identify 76\% of the fossil systems as CC systems. If one uses the concentration parameter alone, we observe that 67\% of the systems are selected as CC systems (though it's quite likely that the $z < 0.05$ fossils will also have $c_{\mathrm{SB}} > 0.075 $, in which case the fraction would go up to 76\%).

\begin{figure*}
\includegraphics[scale = 0.5, angle = 270]{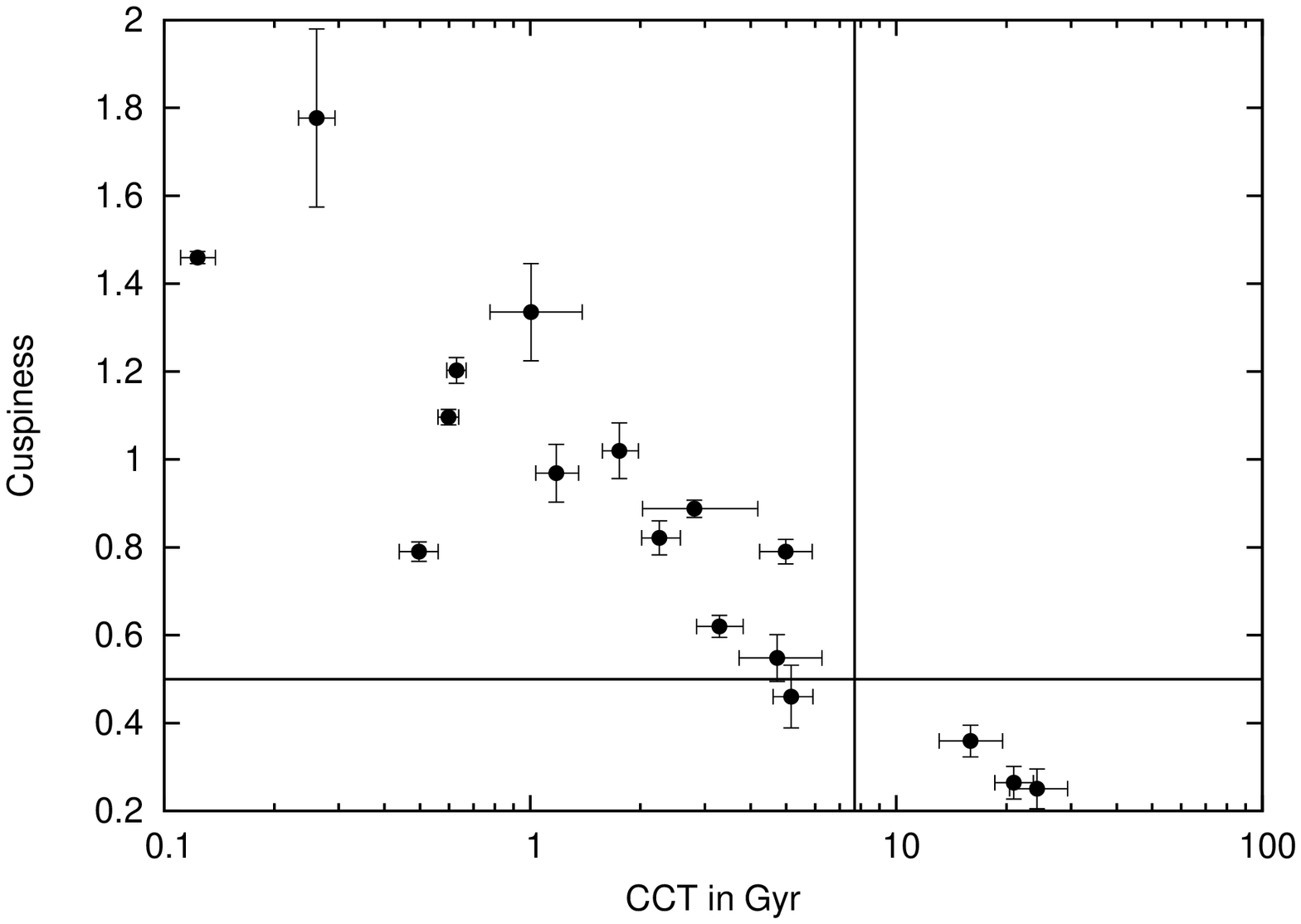}
\includegraphics[scale = 0.5, angle = 270]{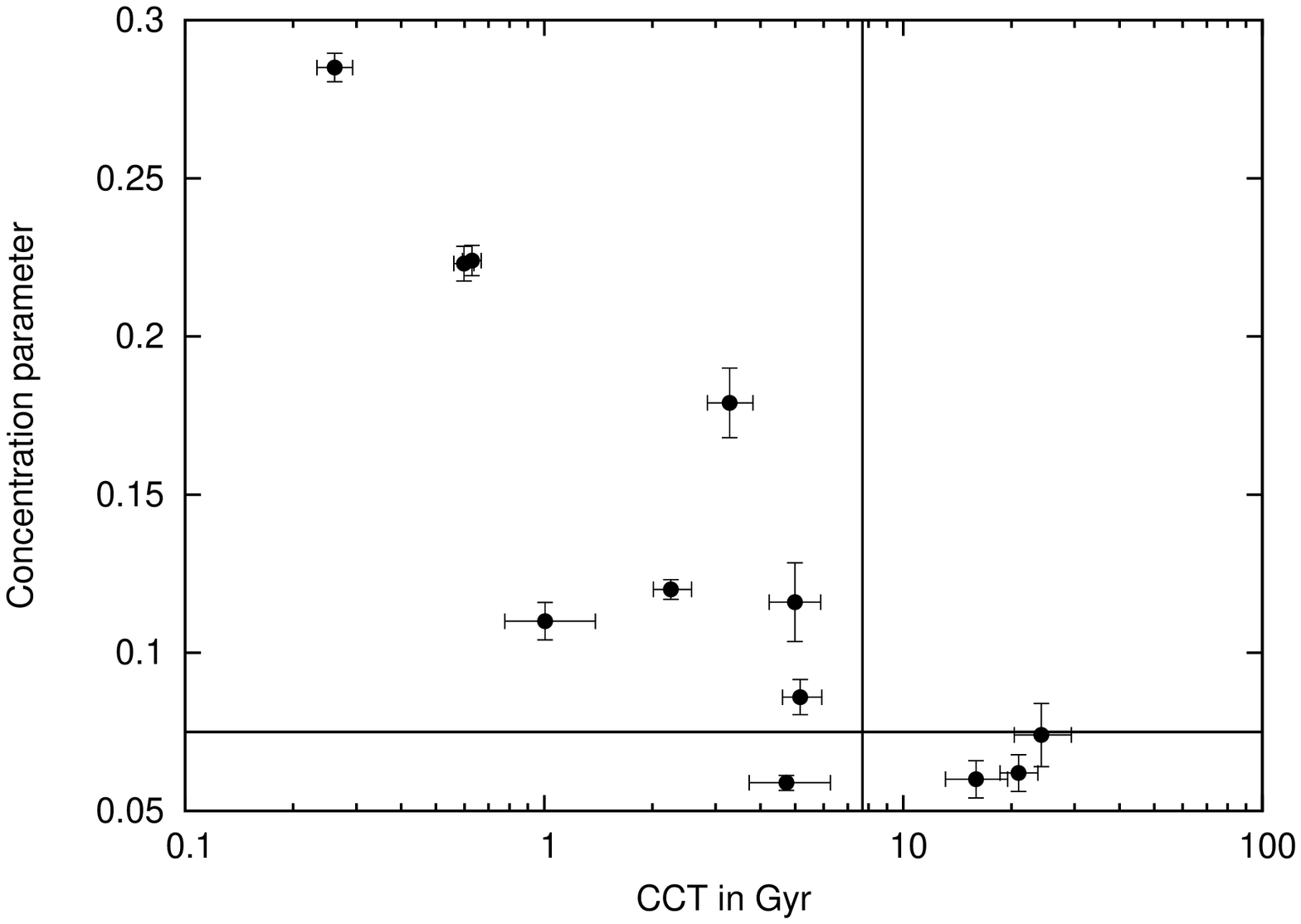}
\caption{Plot of CCT vs.~cuspiness (left) and CCT vs.~concentration parameter (right). The horizontal lines represent the cut for being classified as a cool-core system. The vertical line represents $\mathrm{CCT} = 7.7~\mathrm{Gyr}.$ }\label{CCTalphaconc}
\end{figure*}

Thus, if one were to simply use a single quantity to identify CC groups, conservatively, one would get a lower limit on the fossil CC fraction of 67\% and an upper limit of 82\%, i.e. qualitatively there are indications that a majority of fossil systems are CC systems, albeit as the cooling times indicate, most seem to be WCC systems. Fig.~\ref{CCTalphaconc} shows the relation between the CCT--$\alpha$ and CCT--$c_{\mathrm{SB}}$ and despite the scatter, it is clear that the anti-correlation trend between the quantities is quite strong with Spearman correlation coefficients of -0.91 and -0.80 respectively.

Assuming no significant effect of selection bias on the sample, to the best of our knowledge, this is the first demonstrated result that suggests that most fossils are in fact cool-core systems based on observable diagnostics. The fraction of cool-core fossil systems is marginally higher than that typically reported for objectively selected group (77\%, \citealt{2014A&A...572A..46B}) and cluster samples (72\%, \citealt{2010A&A...513A..37H}), though we do point out that studies have shown that cool-core fractions are generally over-estimated in flux-limited samples of clusters (e.g.~\citealt{2010A&A...513A..37H,2011A&A...526A..79E,2011A&A...532A.133M}). The high cool-core fossils fraction is not totally unexpected as it is generally assumed that fossils are among the most relaxed clusters/groups in the Universe and hence should have pronounced cool-cores with very short CCTs. What is more intriguing is the fact that the WCC fraction is the highest among the three classes, and the population of WCC and NCC fossils combined together (collectively called non-strong cool-cores; NSCCs) \textit{exceed} the SCC fossils population. This is not what one would naively expect from ostensibly relaxed objects such as these and are first indications of an impact of non-gravitational processes on the fossil ICM. 

\subsection{EP-BCG/EP-EWC separation}
We also quantified the dynamical state of the fossil system via two properties, namely the emission peak--emission weighted centre (EWC) separation and the central elliptical galaxy--EP separation, i.e. the BCG--EP separation. These are reasonable indicators of whether the system is relaxed or not, as large values for these numbers would indicate a disturbed system (e.g.~\citealt{2010A&A...513A..37H}).

The EWC was determined iteratively in a point source subtracted, exposure-corrected image by starting off from the EP as was done in \cite{2006A&A...453..433H}. The positions of the BCG were taken from literature and was confirmed from the NASA extragalactic database (NED)\footnote{\url{http://ned.ipac.caltech.edu/}}.

\begin{figure*}
 \includegraphics[scale=0.5,angle=270]{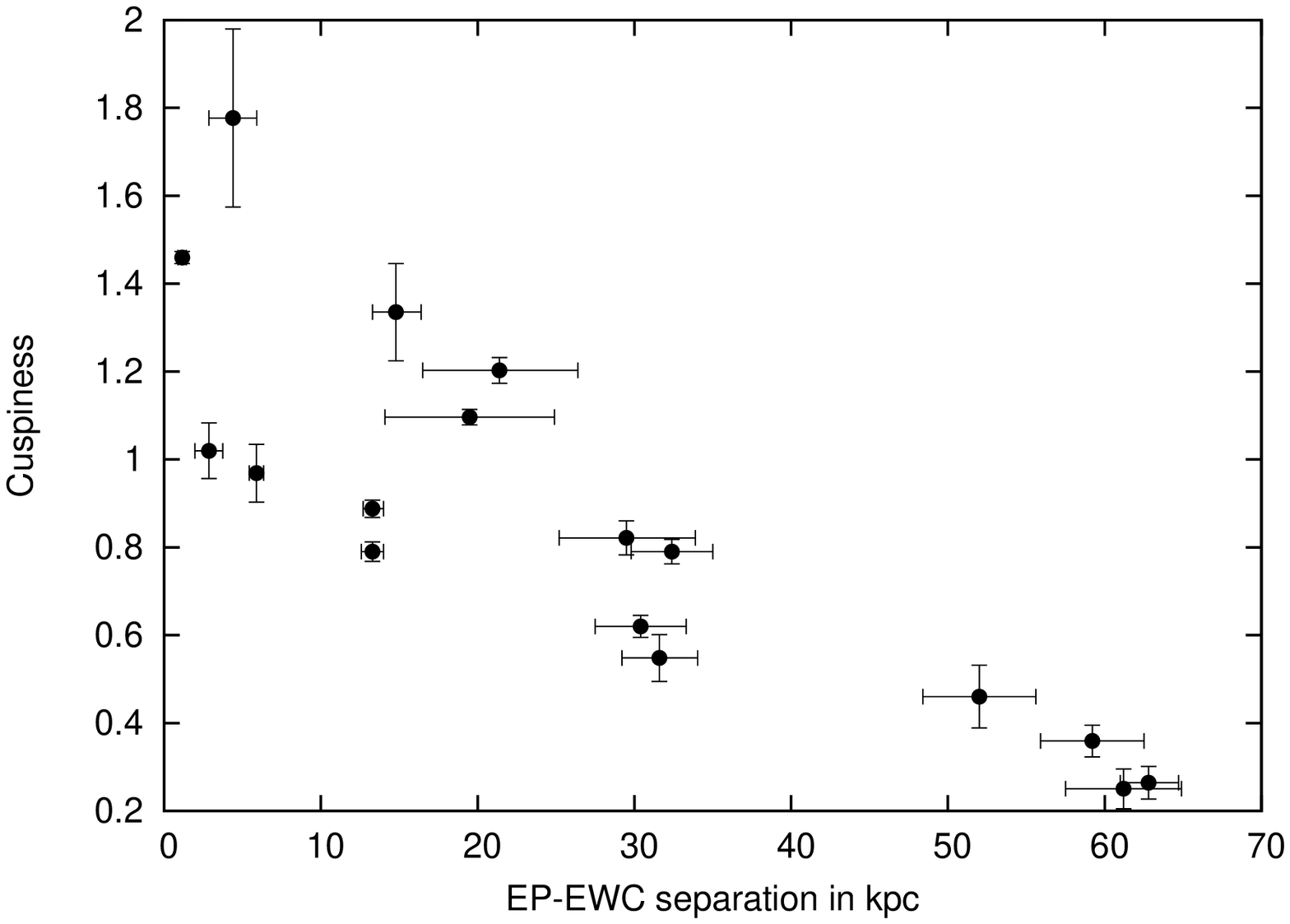}
 \includegraphics[scale=0.5,angle=270]{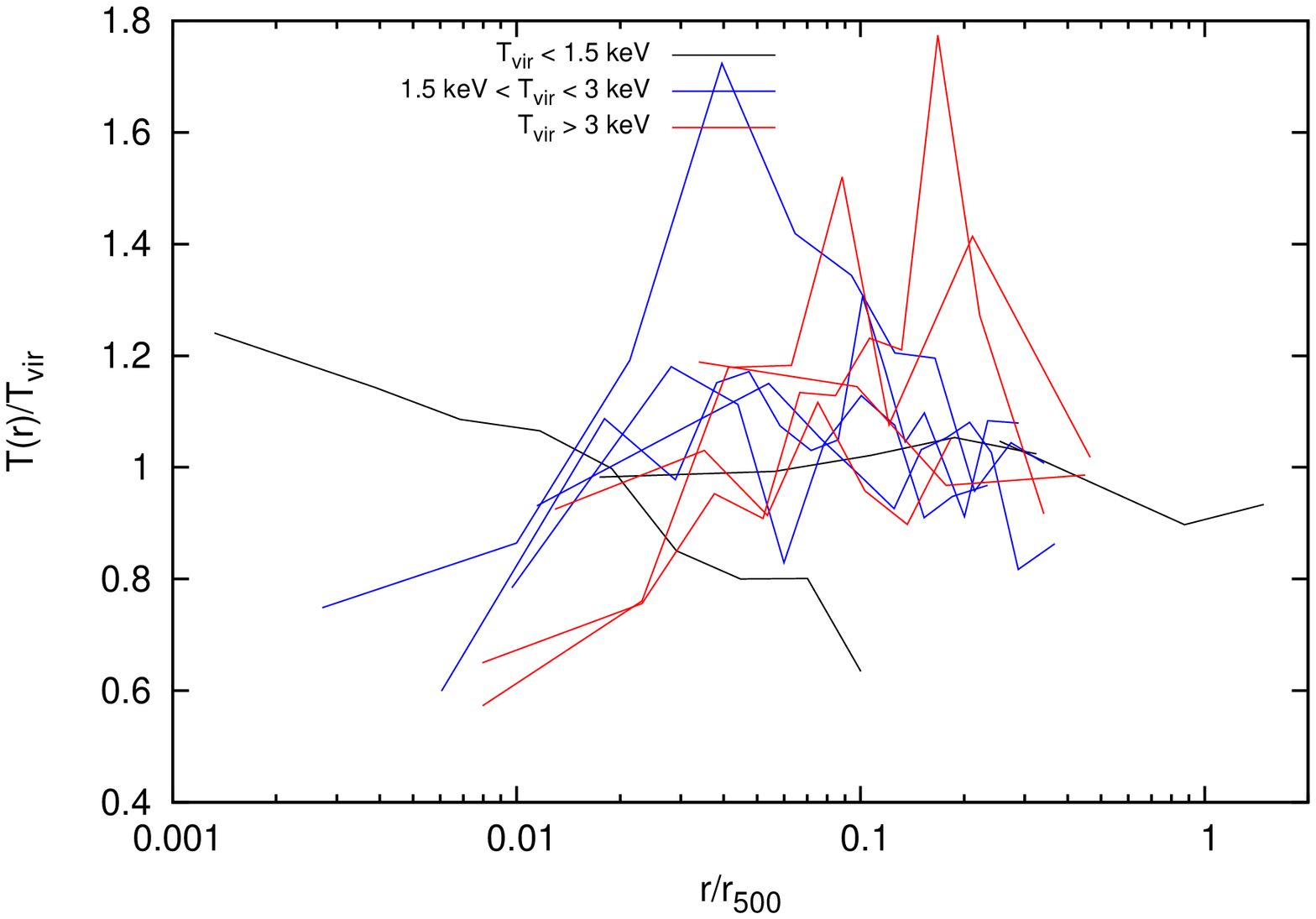}
\caption{\textit{Left:}Plot of EP-EWC separation and cuspiness. \textit{Right:} Scaled temperature profiles colour-coded as a function of virial temperature.}\label{EPEWC_cusp}
\end{figure*}

Table~\ref{distkpc} shows the EP--EWC and EP--BCG separation. All fossils have their BCG located within 50 kpc of the X-ray peak. The EP--EWC peak shows a wider range of values with four fossils (24\%) having a separation $\geq$ 50 kpc. There is an anti-correlation trend of the EP--EWC separation with the Cuspiness with a Spearman correlation coefficient of -0.86, i.e. cool-core fossils have a smaller separation and vice-versa (Fig.~\ref{EPEWC_cusp}, left)

Despite most fossils having an EP--EWC separation less than 50 kpc, again such a large variation in EP--EWC values is not naively expected for ostensibly relaxed systems such as fossils.

\begin{table*}
\centering
\caption{EP--BCG, EP--EWC separation. The columns are (1) name of system, (2) separation between BCG and EP in kpc, (3) separation between EP and EWC in kpc. The uncertainty on these quantities correspond to $1^{''}$ which in turn corresponds to 0.27 kpc for the nearest fossil, and 5.41 kpc for the furthest one.}
 \begin{tabular}{| c | c | c |}
  \hline \hline
  Name & BCG-EP (kpc) & EP-EWC (kpc)\\ \hline
  NGC~6482 &0.212&1.16 \\
  NGC~1132 &0.785 &5.90 \\
  RX~J0454.8-1806 &3.22 &13.3 \\
  ESO~306-G 017 &0.842 &13.3 \\
  UGC~842 &0.960 &2.87 \\
  RX~J1331.5+1108 &1.71 &14.8 \\
  cl1159p5531 &3.77 &4.40 \\
  cl2220m5228 &3.71 &62.8 \\
  cl1416p2315 &0.964 &31.6 \\
  cl0245p0936 & 2.78 & 32.4\\
  cl1340p4017 &3.46 &30.4  \\
  RX~J2247.4+0337 &2.93 &59.2  \\
  RX~J0825.9+0415 &3.58 &52.0 \\
  RX~J1256.0+2556 &6.6 &61.2  \\
  RX~J0801+3603 &1.29 &29.5 \\
  RX~J1115.9+0130 &7.88 &21.4 \\
  RX~J0159.8-0850 &4.31 &19.5 \\ \hline
\hline
 \end{tabular}
\label{distkpc}
\end{table*}

\subsection{Temperature Profiles}
Figure~\ref{EPEWC_cusp}, right and appendix~\ref{Tempprofiles} shows the scaled temperature profiles for the fossil systems for which we could obtain one (12 objects). There's a great degree of variation in the profile shapes with temperature drops and flat shapes for the fossil systems and we fail to pinpoint a ``typical'' temperature profile for the fossils. The high redshift, high temperature systems show temperature profiles that are expected for a SCC/WCC cluster, i.e. indications of a temperature drop for the SCC objects and a relatively flat profile for the WCC objects \citep{2010A&A...513A..37H}. NGC 6482 shows a profile that rises all the way to the centre which has been reported thoroughly in previous studies of the object \citep{2004MNRAS.349.1240K}. NGC 1132 shows an almost a flat profile with barely a decline towards the centre. ESO306, RX~J0454.8-1806, and UGC00842 lacks a smooth temperature decline towards the centre which while potentially indicative of a cool-core, shows that it's a truncated one. One possibility for the shape of such a temperature profile could be emission from the BCG, which we discuss in Sec.~\ref{coronasec}. A deprojection analysis (Sec.~\ref{deprojsec}) for the $z < 0.05$ systems indicates the imprint of a heating process such as AGN. Indeed, only cl1159p5531 shows the ``classic'', peaky temperature profile observed for low-temperature groups with a SCC (e.g.~\citealt{2009ApJ...693.1142S,Eckmiller}) versus that for SCC clusters which show flattening at intermediate radii. (e.g.~\citealt{2006ApJ...640..691V}).

\subsection{Potential emission from the BCG}\label{coronasec}
As shown in Table~\ref{distkpc} most fossils have a BCG which is very close to or on the X-ray peak. The source detection tool \verb+wavdetect+ manages to detect sources which seem to be coincident with the BCG in most cases and we explored whether part of the emission in the core regions comes from the BCG itself, i.e. if they are coronae which consists of gas of stellar origin. This could in principle also explain the truncated ``cool-cores'' seen in objects such as UGC00842, ESO306, and RX~J0454.8-1806. To study this emission, we focused our attention only on the fossil systems with $z<~0.1$, as it becomes harder to resolve potential coronae for high redshift objects. Using the regions generated by \verb+wavdetect+ we extracted spectra which were fit to an absorbed APEC model. The regions used for the spectral extraction are of the order of a few kpc, which is the typical size of a corona. We kept the temperature, abundance, and normalisation as the free parameters. The best-fit values were used to determine the X-ray luminosity using \textit{Xspec} in the 0.5--2.0 keV band, which was then compared with the ROSAT luminosity of the fossil, to determine the strength of this ``corona'' emission.

Our estimated temperatures from the spectral fit is in most cases much higher ($\geq~1$ keV) than that is typically observed for the corona class. Moreover, the X-ray luminosity of the emission is in the order of $10^{41}$ erg/s which is an order of magnitude higher than what is typically observed for coronae \citep{2007ApJ...657..197S} and can account for as much as $17\%$ of the total fossil ROSAT luminosity. Lastly, the abundances are much lower (median of $0.28$) than what has been reported for true coronae ($0.8$, \citealt{2007ApJ...657..197S}) which gives indications that the detected emission might not be completely of stellar origin, though we do point out that there is evidence from literature for low-metallicity coronae as well (e.g.~\citealt{2014MNRAS.439.1182S}). One strong possibility is that the stellar content has mixed with the surrounding dense ICM. Table~\ref{corona} shows the properties of the emission associated with the BCG. 

\begin{table*}
\centering
 \caption{X-ray emission coincident with the BCG for $z\leq 0.1$ fossils. Columns are (1) Name of group, (2) Temperature, (3) metallicity, (4) X-ray luminosity from Xspec, (5) ``Corona'' luminosity as percentage of ROSAT luminosity }\label{corona}
 \begin{tabular}{|c|c|c|c|c|}
 \hline \hline & & & &\\
 Name & $T$ & $Z_{\odot}$ & $L_{\mathrm{X}} (0.5-2.0 \mathrm{keV}) (10^{41}~\mathrm{erg/s})$ & $\%$ of ROSAT group emission \\ \hline & & & & \\
 NGC~6482 &$0.95^{+0.022}_{-0.022} $ &$0.28^{+0.08}_{-0.06} $ &$1.05$ &17\% \\ 
 NGC~1132 & $0.82^{+0.06}_{-0.08}$ &$0.11^{+0.05}_{-0.04} $ &$1.78$ &3.8\% \\ 
 RX~J0454.8-1806 &$1.04^{+0.16}_{-0.08}$ &$0.40^{+0.55}_{-0.19} $ &$1.06$ &1.2\% \\ 
 ESO~306-G 017 &$1.30^{+0.04}_{-0.05} $ &$0.29^{+0.12}_{-0.09} $ &$6.46 $ &2.4\% \\ 
 UGC~842 &$1.39^{+0.13}_{-0.07}$ &$0.21^{+0.11}_{-0.07} $ &$1.78$ &7.1\% \\ 
RX~J1331.5+1108 &$0.89^{+0.063}_{-0.034}$ &$0.15^{+0.54}_{-0.10} $ &$1.84 $&2.9\%  \\ 
cl1159p5531 &$1.28^{+0.02}_{-0.02}$ &$0.38^{+0.06}_{-0.05} $ &$17.2$ &15\% \\
 
 \hline
 \end{tabular}

\end{table*}

\subsection{Deprojection analysis of $z<0.05$ fossils}\label{deprojsec}

In order to explore thermodynamic properties of fossil systems in detail, we focused in particular on the low redshift ($z<0.05$) fossils (five in total, namely, NGC 6482, NGC 1132, ESO306, RXJ~0454.8-1806, UGC 00842) and performed a deprojection analysis. To do this, we adopted the MBPROJ code \citep{2014MNRAS.444.1497S} which is based on the surface brightness deprojection method of \cite{1981ApJ...248...47F} and is used to estimate the ICM thermodynamical properties from the surface brightness profiles. This method potentially has the advantage of resolving the extreme inner regions of the galaxy group which cannot be probed with traditional spectral analysis in the absence of very high quality data. Additionally, this is a good opportunity to test the code on low temperature groups to see how well the results agree with spectral analysis and test the usability of the code for situations where spectral analysis would not be possible (e.g. survey data).

For purposes of brevity we refrain from describing the code in detail and guide the reader to the aforementioned paper. The relevant information for this particular study are as follows:

\begin{itemize}
 \item Surface brightness profiles were extracted in the 0.5-1, 1-1.5, 1.5-2.5, 2.5-7 keV bands.
 \item We focused only on the ``null potential'' case for this particular study. This particular case does not assume hydrostatic equilibrium and essentially corresponds to low resolution spectral fitting. The advantage of this case is that it prevents the thermodynamic profiles from being biased due to assumption of hydrostatic equilibrium. The null potential case can be used to estimate thermodynamic properties with and without linear interpolation of the temperatures between the bins. The non-interpolation case assumes the temperature is the same in groups of three density bins. For the scientific discussion here, we focused only on the profiles which do not involve linear interpolation of the temperatures.
 \item The abundance is assumed to be 0.3$Z_{\odot}$ throughout.
\end{itemize}

\begin{figure*}
\includegraphics[scale=0.50,angle=270]{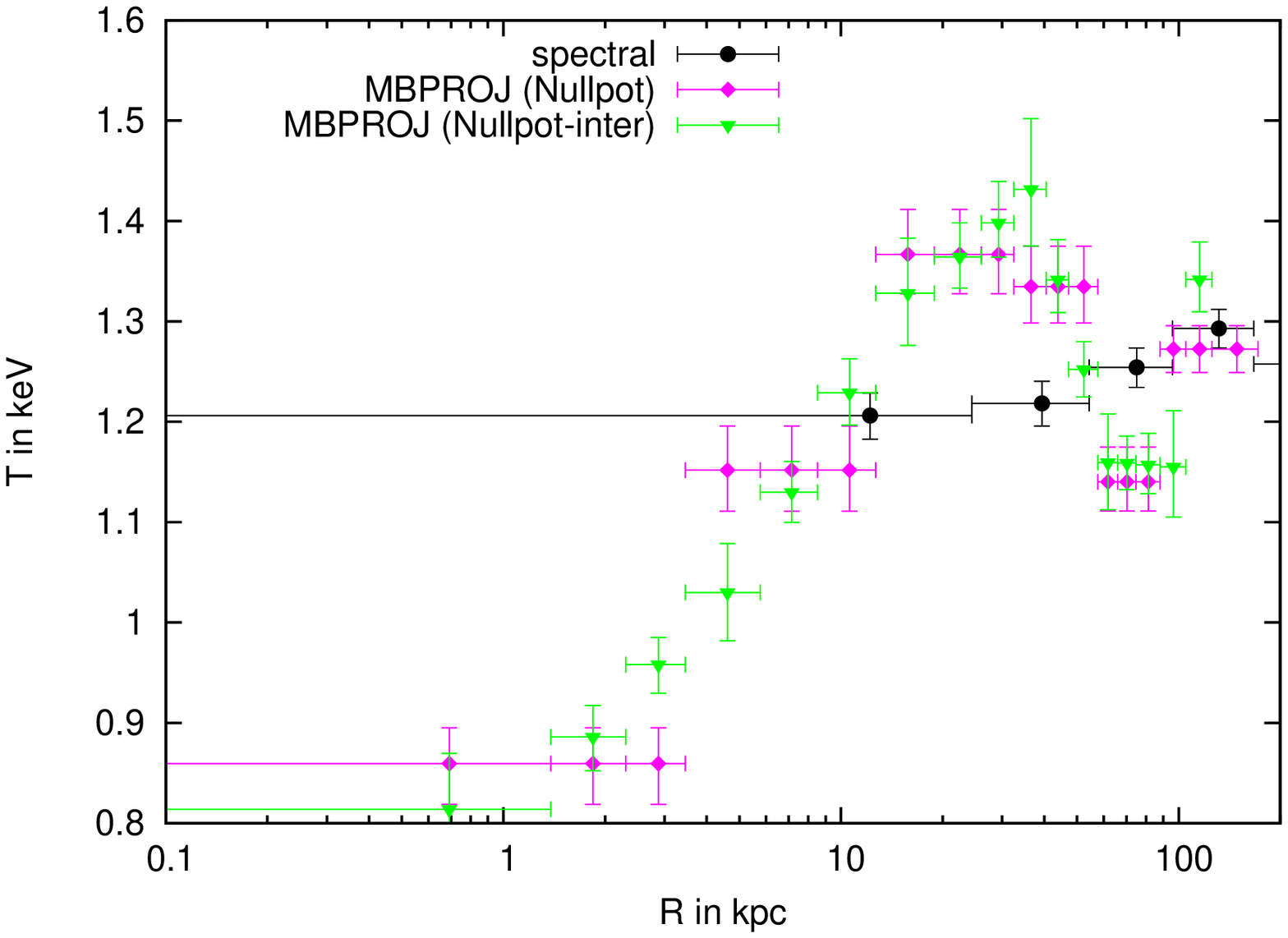}
\includegraphics[scale=0.50,angle=270]{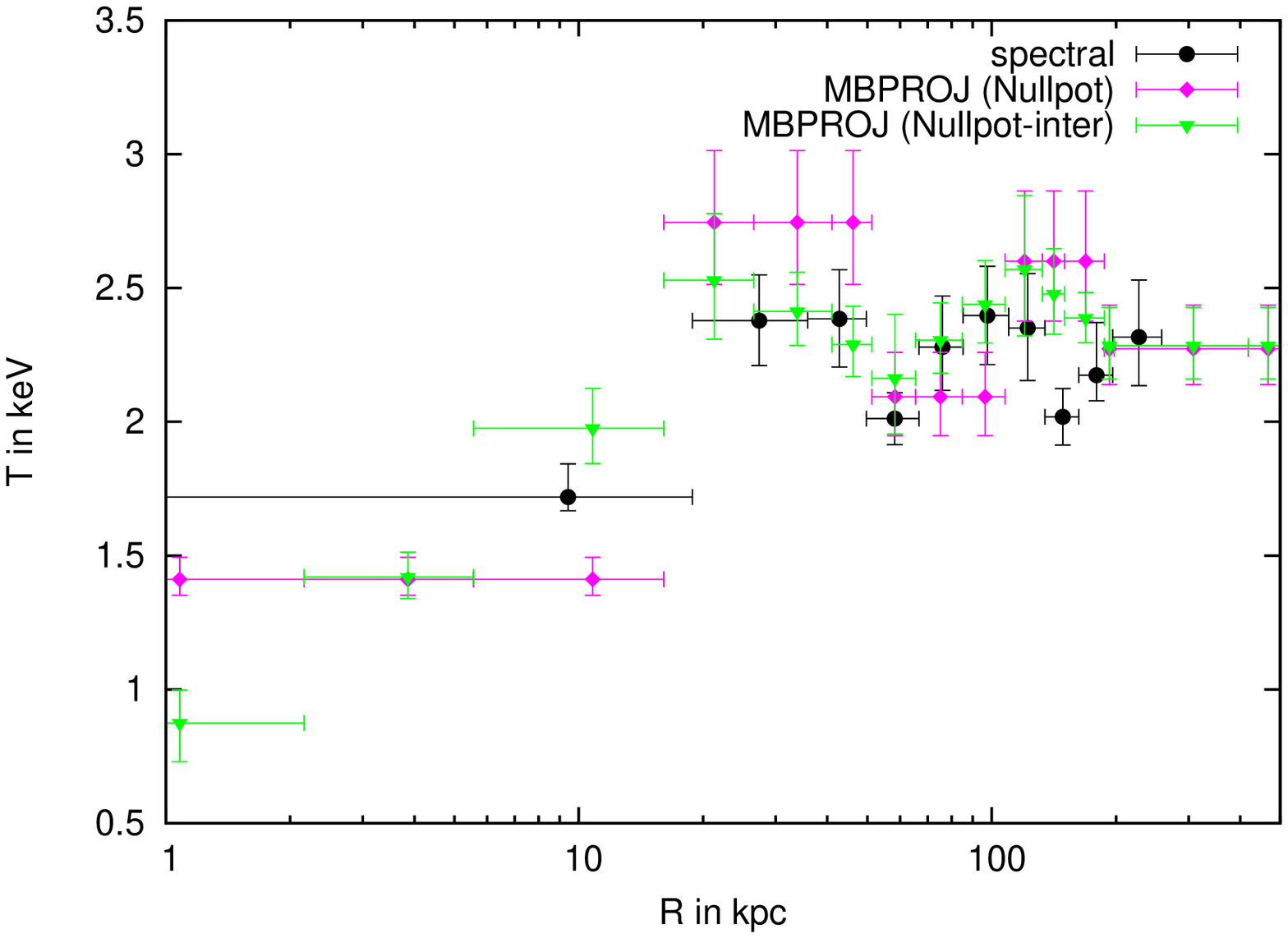}
\includegraphics[scale=0.50,angle=270]{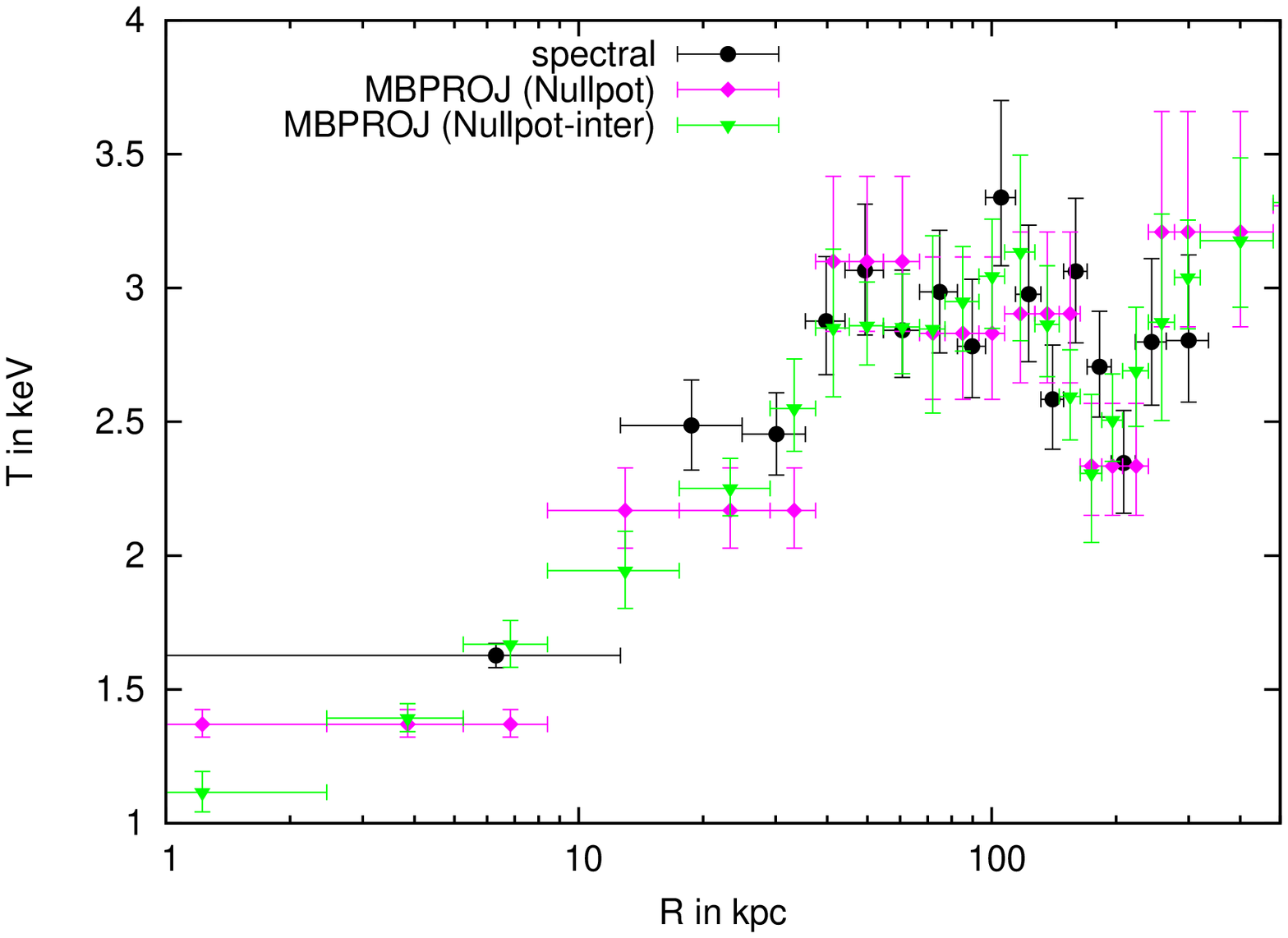}
 \includegraphics[scale=0.50,angle=270]{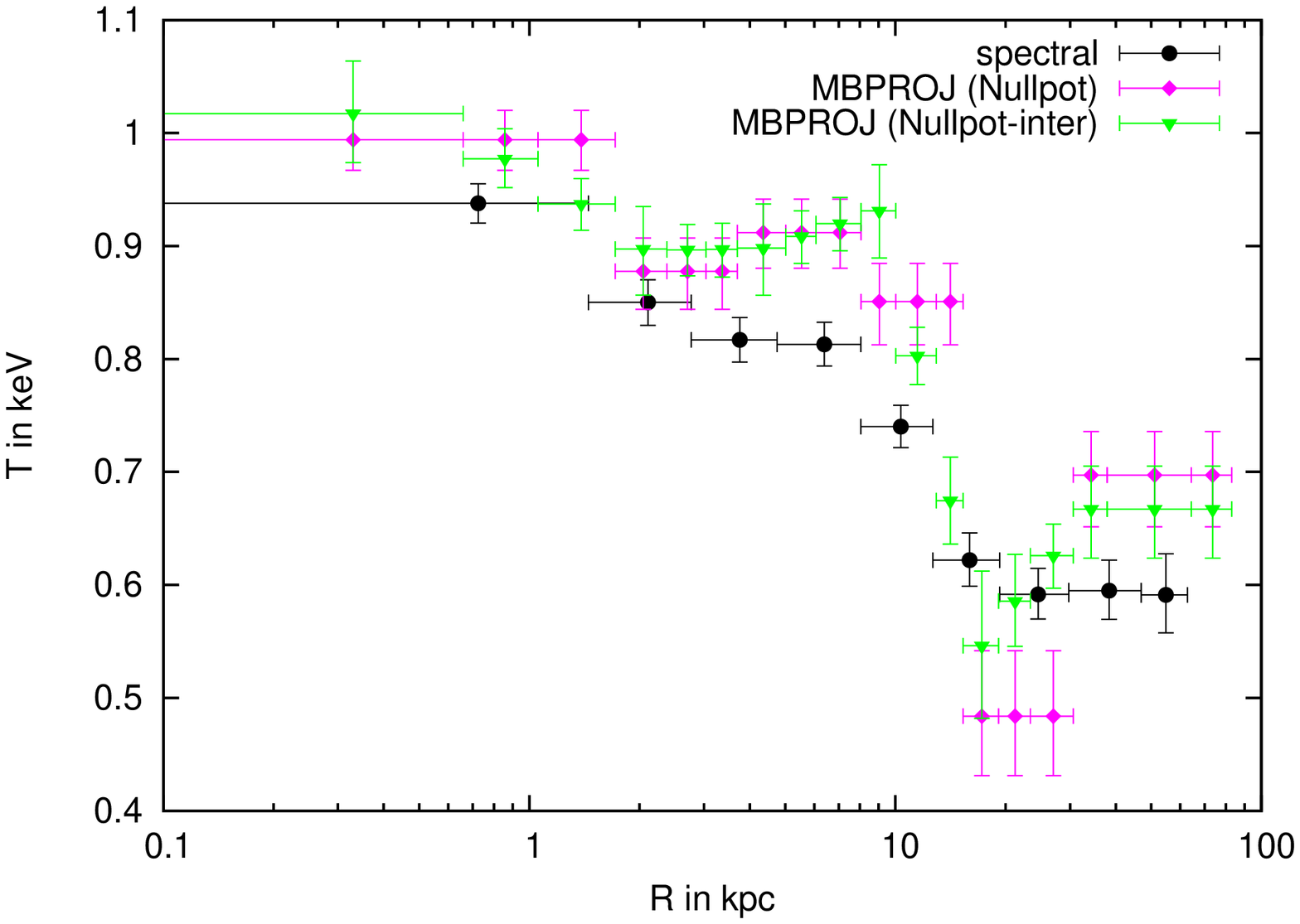}
\includegraphics[scale=0.50,angle=270]{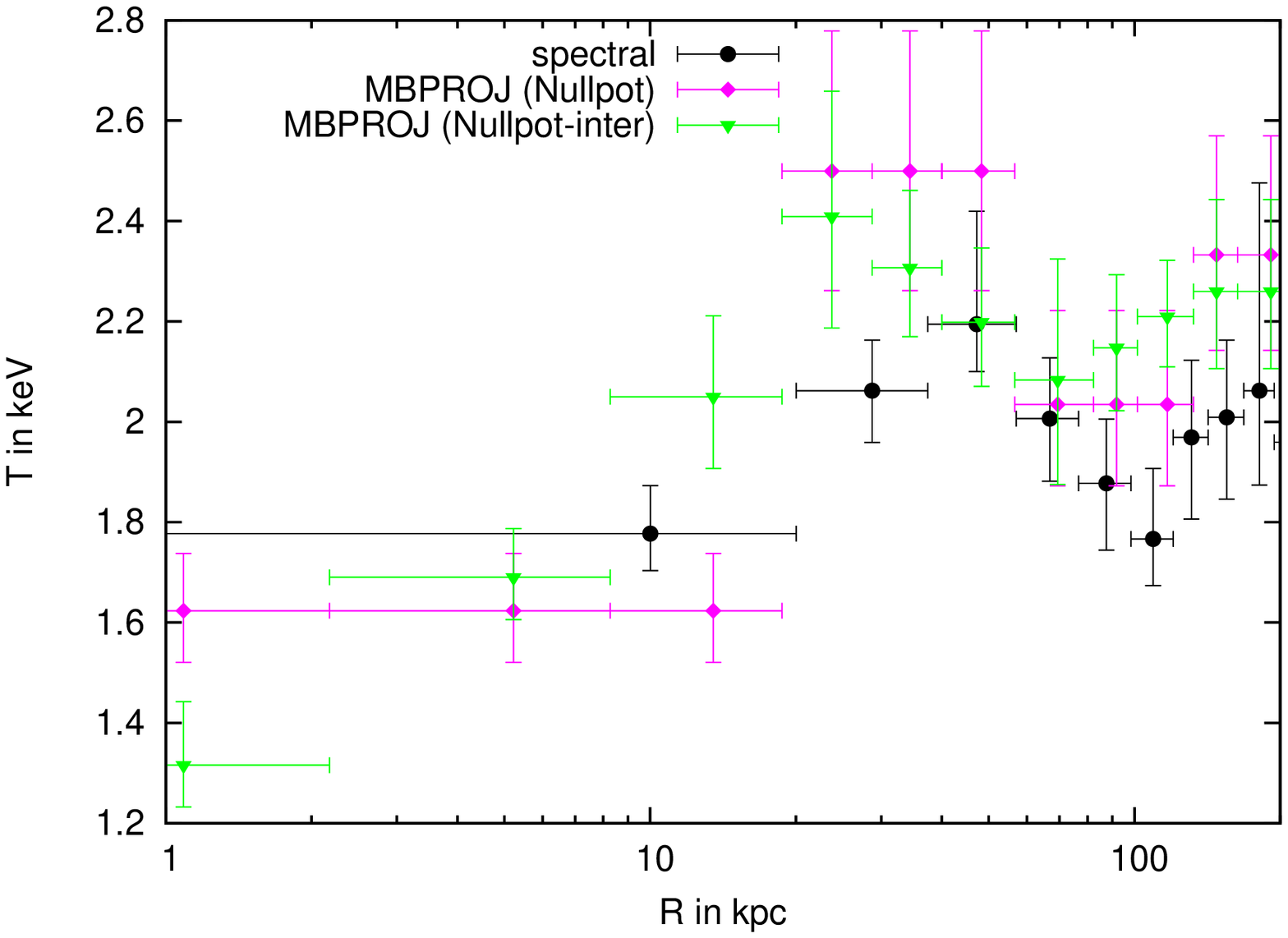}
\caption{Temperature profiles of the fossil systems, from spectral analysis and MBPROJ. \textit{Clockwise from top left}: NGC 1132, RXJ~0454.8-1806, NGC 6482, ESO306, UGC 00842. The spectral temperature profiles were also determined for an abundance of 0.3$Z_{\odot}$.}\label{TprofsMBPROJ}
\end{figure*}

Figure~\ref{TprofsMBPROJ} shows a comparison of the temperature profiles from the spectral analysis and MBPROJ. The code does a good job of recovering the shape of the temperature profiles though some deviations are seen for NGC 1132 and NGC 6482 which could be due to the extremely low temperatures of these systems ($1.2$ and $0.75$ keV respectively), and possible projection effects. For the hotter groups, the agreement is much better. Overall though, MBPROJ seems to be a competitive code even on the group regime and would be particularly useful for eROSITA data where most systems would be galaxy groups and would lack sufficient counts for a direct spectral temperature estimate \citep{2014A&A...567A..65B}. 

\begin{figure*}
\centering
 \includegraphics[scale = 0.50,angle=270]{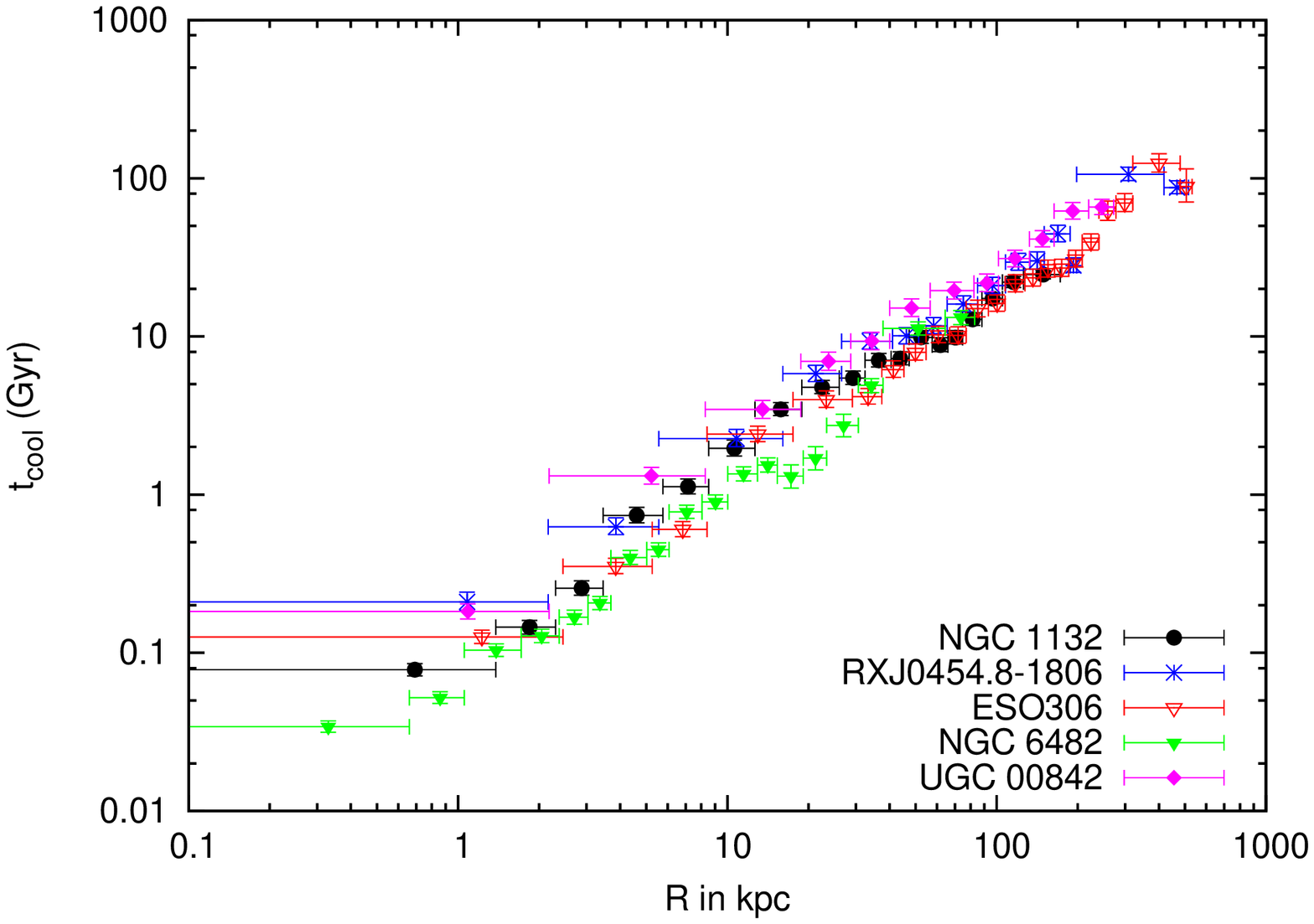}
\includegraphics[scale=0.50,angle=270]{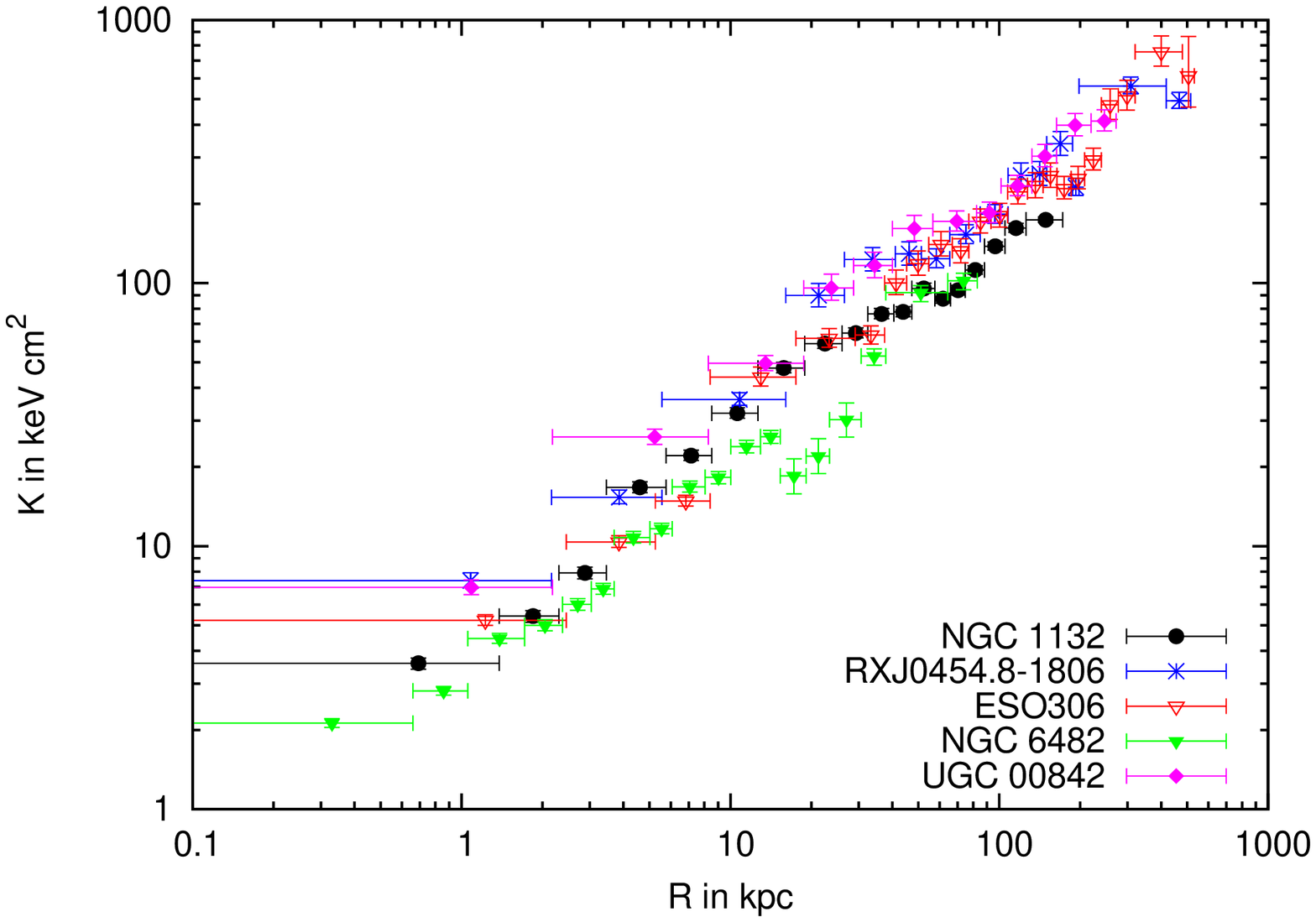}
\includegraphics[scale=0.50,angle=270]{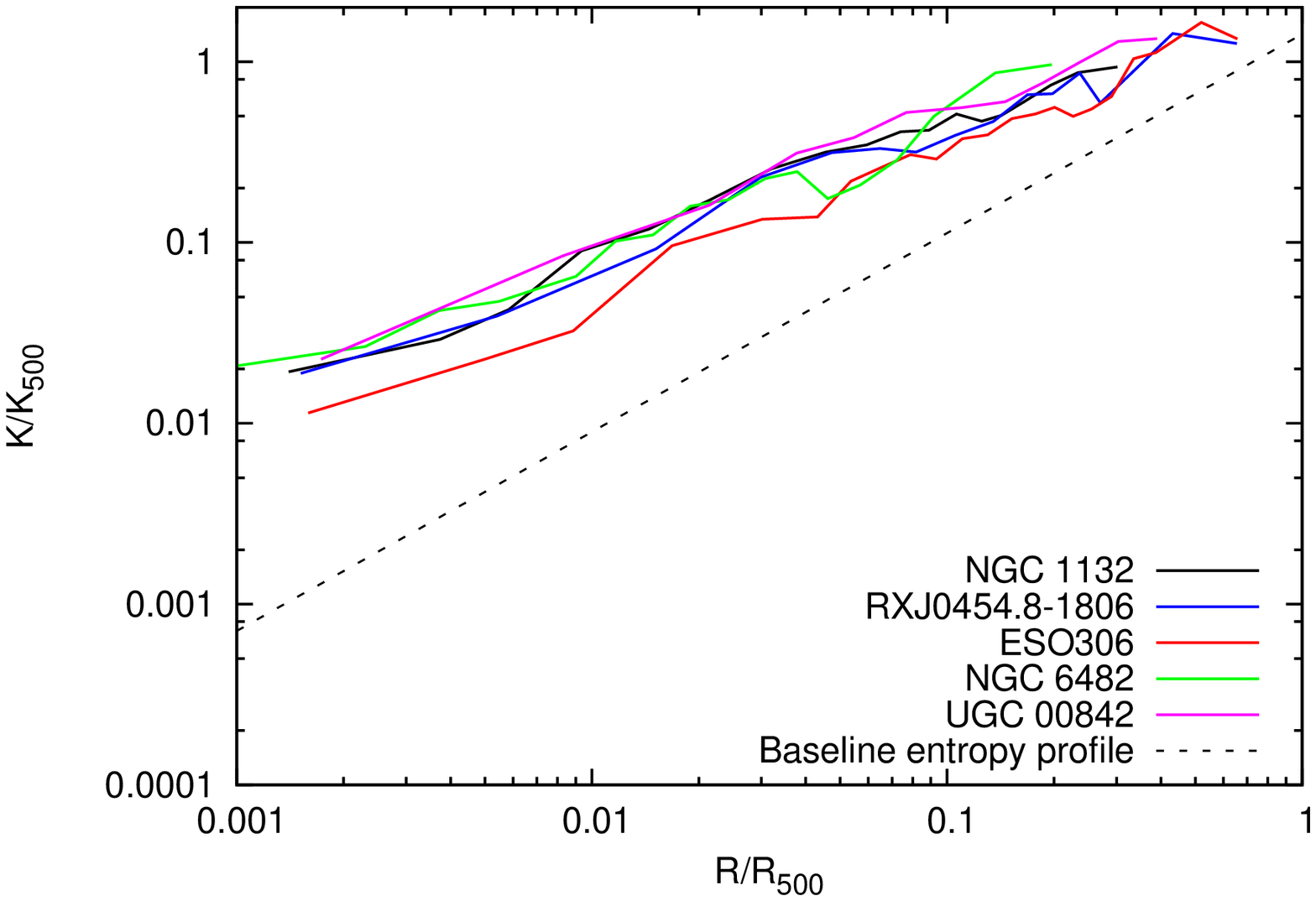}
\caption{\textit{Top:} Cooling time profiles of $z < 0.05$ fossils. \textit{Middle:} Entropy profiles of $z < 0.05$ fossils. Both quantities are obtained from the non-interpolated temperatures estimated from the MBPROJ code. \textit{Bottom:} Scaled entropy profiles. Dashed line represents the baseline entropy profile from \cite{2010A&A...511A..85P}.}\label{tcoolentropy}
\end{figure*}

Figure~\ref{tcoolentropy} shows the cooling time and the entropy profiles of the fossil systems. The cooling time profiles can be well-described by a powerlaw with an index varying from $1.06\pm0.03$ to $1.20\pm0.05$ for the five systems. \cite{2010A&A...513A..37H} define the cooling radius as the radius at which $t_{\mathrm{cool}} = 7.7$ Gyr. For these five fossils, namely, NGC 1132, RX~J0454.8-1806, ESO306, NGC 6482, and UGC 00842, this corresponds to 43 kpc, 36 kpc, 51 kpc, 54 kpc, and 30 kpc respectively. However, as shown in Fig.~\ref{TprofsMBPROJ}, the region with the cool gas (if present) is generally much smaller than the cooling radius. MBPROJ detects cool gas out to 3 kpc for NGC 1132, 12 kpc for RX~J0454.8-1806, 9 kpc for ESO306, 0 kpc for NGC 6482, and 19 kpc for UGC 00842. Thus, all these systems seem to lack a group-sized cool-core and the region of cool gas is not commensurate to the region of short cooling times. As argued above, it seems unlikely that the cool gas component is emission purely associated with the BCG.  

The derived entropy profiles can also be well-described by a powerlaw of the form $K(r) = K_{100} \times (r/100~\mathrm{kpc})^{m}$ as has been used in studies such as e.g.~\cite{2009ApJS..182...12C} and \cite{2014MNRAS.438.2341P}. When the profile is fit with the above function, a powerlaw index that is substantially lower than 1.1 is obtained; which is what is expected for pure gravitational processes (e.g.~\citealt{2001ApJ...546...63T}). \cite{2009ApJS..182...12C} also fit their entropy profiles with a powerlaw model with an additional component for a non-zero core entropy, i.e.~$K(r) = K_{0} + K_{100} \times (r/100~\mathrm{kpc})^{m}$. We also added this component to see if it has an effect on the fitted entropy profile. As shown in Table~\ref{entable}, the index is still much lower than 1.1, and in some cases, $K_{0}$ takes unphysical negative values. Qualitatively, this agrees with the results of \cite{2014MNRAS.438.2341P} who showed that the entropy profiles of a sample of galaxy groups and clusters can be sufficiently described by only a powerlaw without indications of any core-flattening. We also compared the scaled entropy profiles to the baseline entropy profile of \cite{2010A&A...511A..85P}, which is defined as:
\begin{equation}
 K(R)/K_{500} = 1.42 (R/R_{500})^{1.1}
\end{equation}
where $K_{500}$ is as defined in \cite{2010A&A...511A..85P} and is dependent on $M_{500}$. To estimate $M_{500}$ we used the $M_{500}-T$ scaling relation from \cite{2005A&A...441..893A}. As is evident in Fig.~\ref{tcoolentropy} there is a significant entropy excess over the baseline entropy profile for the fossils within the radial ranges probed here, and the profile is considerably flatter as compared to the baseline profile. The shallow slopes, and the elevated entropies over a substantially large radial range is indicative of a strong influence of non-gravitational processes on the properties of the ICM, which we discuss in greater detail in Sec.~\ref{Disc}.

\begin{table*}
\centering
\caption{Best-fit results for Entropy profiles of $z < 0.05$ fossils, without and with core component. Columns are (1) Name of group, (2) Slope, (3) Normalisation, (4) core entropy }\label{entable}
 \begin{tabular}{|c|c|c|c|}
 \hline \hline
  Name & $m$ & normalisation ($K_{100}, \mathrm{keV~cm}^{2}$) & $K_{0}~\mathrm{keV~cm}^{2}$ \\ \hline 
  NGC~6482 &$0.74\pm0.04$ &$103.3\pm13.12$ &--- \\
  NGC~1132 &$0.75\pm0.02$ &$138.9\pm6.08$ &--- \\
  RX~J0454.8-1806 &$0.72\pm0.02$ &$184.2\pm9.83 $ &--- \\
  ESO~306-G 017 &$0.83\pm0.02$ &$170.1\pm9.00$ &--- \\
  UGC~842 &$0.75\pm0.02$ &$224.49\pm8.45 $ &--- \\ 
  All 5 groups & $0.80\pm0.01$ & $146.1\pm5.41$ &--- \\ \hline \hline

  NGC~6482 &$0.91\pm0.05$ &$139.7\pm16.3 $ &$1.24\pm0.29$ \\
  NGC~1132 &$0.70\pm0.04 $ &$138.9\pm5.96$ &$-1.38\pm1.30$ \\
  RX~J0454.8-1806 &$0.72\pm0.05$ &$184.5\pm11.1$ &$-0.14\pm2.38$ \\
  ESO~306-G 017 &$0.90\pm0.04 $ &$168.9\pm8.25$&$1.83\pm0.84$ \\
  UGC~842 &$0.70\pm0.03$ &$225.1\pm7.72$ &$-2.43\pm1.53$ \\ 
  All 5 groups & $0.84\pm0.02$ & $147.5\pm5.38$ &$0.74\pm0.36 $ \\ \hline 
 \end{tabular}

\end{table*}

\subsection{$L_{\mathrm{X}}-T$ relation for 400d fossil systems}\label{LTsec}
The $L_{\mathrm{X}}-T$ relation is an interesting scaling relation which can give insights into the baryonic physics at play in the ICM as the physics affects both quantities, $T$ and $L_{\mathrm{X}}$. For example, clusters with cool-cores/non-cool cores have $L_{\mathrm{X}}-T$ relations with different slopes and normalisations, and excising the core regions result in a lowered intrinsic scatter (e.g.~\citealt{2009A&A...498..361P,2011A&A...532A.133M,2012MNRAS.421.1583M}). Note that along with cool-cores and possibly AGN feedback, selection effects can also play a significant role in determining ``true'' scaling relations as has been shown by several authors before (e.g.~\citealt{2002A&A...383..773I,2006ApJ...648..956S,2007MNRAS.382.1289P,2010MNRAS.406.1773M,2011A&A...532A.133M, 2015A&A...573A.118L, 2015A&A...573A..75B}; also see \citealt{2013SSRv..177..247G} section 7, for a review). Thus, simply using a sample of clusters/groups to construct scaling relations without accounting for potential selection effects would not result in an accurate description of the underlying properties of the objects. In this work, there are seven fossil systems from the 400d catalogue (out of 12). The 400d catalogue has fairly well-defined selection criteria, making it relatively easy to account for selection effects, and study if the $L_{\mathrm{X}}-T$ scaling relation for fossils show features that are different to non-fossils. To compare the results here, we used the groups in the \cite{2015A&A...573A..75B} study. To maintain consistency with that study, we used the best-fit virial temperatures and abundances of the fossils, to convert the catalogue ROSAT (0.5--2.0 keV band) luminosities into bolometric luminosities (0.01--40 keV band). 

To determine the scaling relation, we used the BCES (Y$\lvert$X) code by \cite{1996ApJ...470..706A} and the fits were performed in Log space using the fitting function:

\begin{equation}
 \left ( \frac{L_{\mathrm{X}} (0.01-40 \mathrm{keV})}{0.5\times 10^{44}~\mathrm{erg~s}^{-1}} \right ) = c\times\left ( \frac{T}{3~\mathrm{keV}} \right )^{m}
\end{equation}

\begin{figure*}
 \includegraphics[scale=0.50,angle=270]{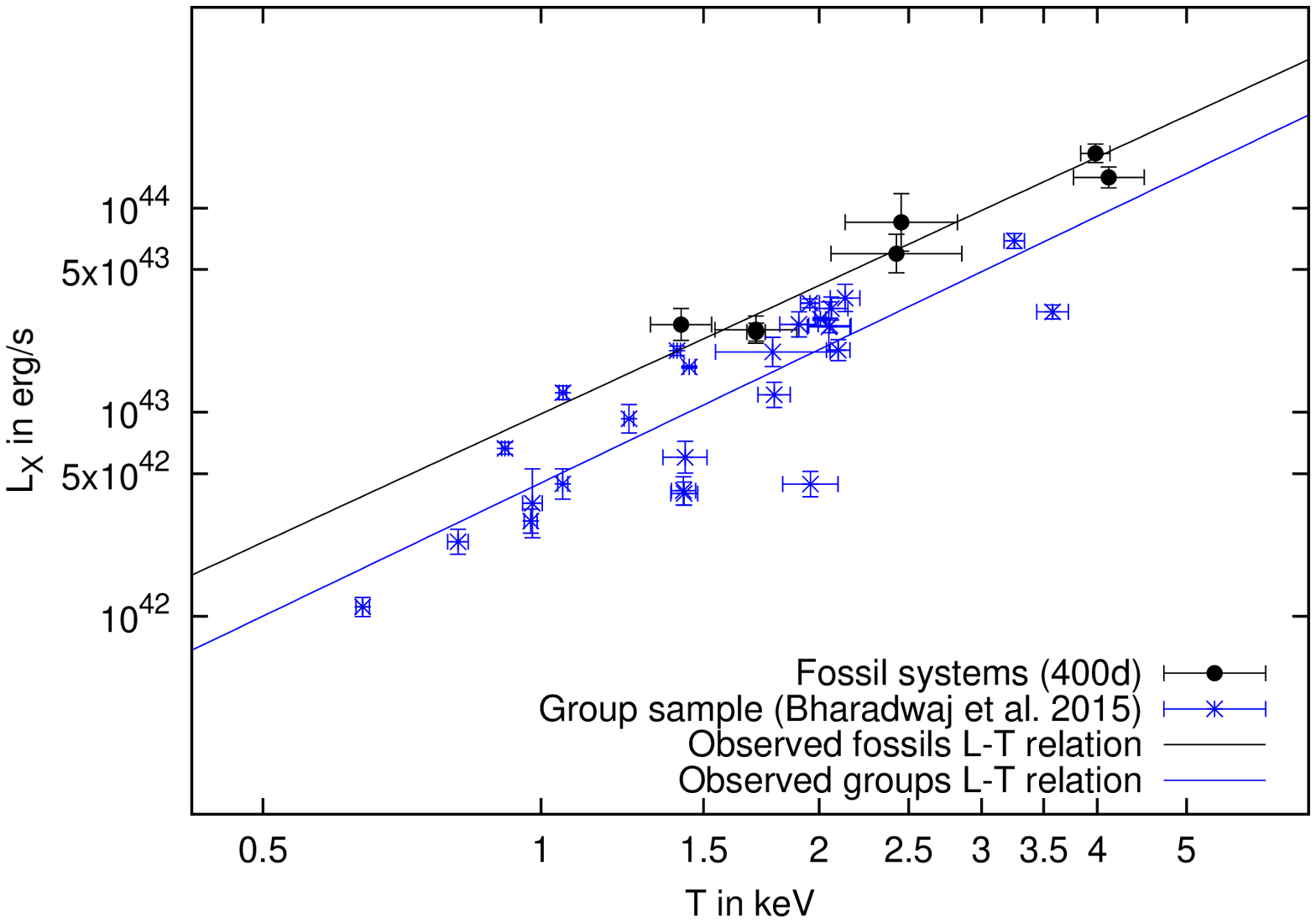}
\includegraphics[scale=0.50,angle=270]{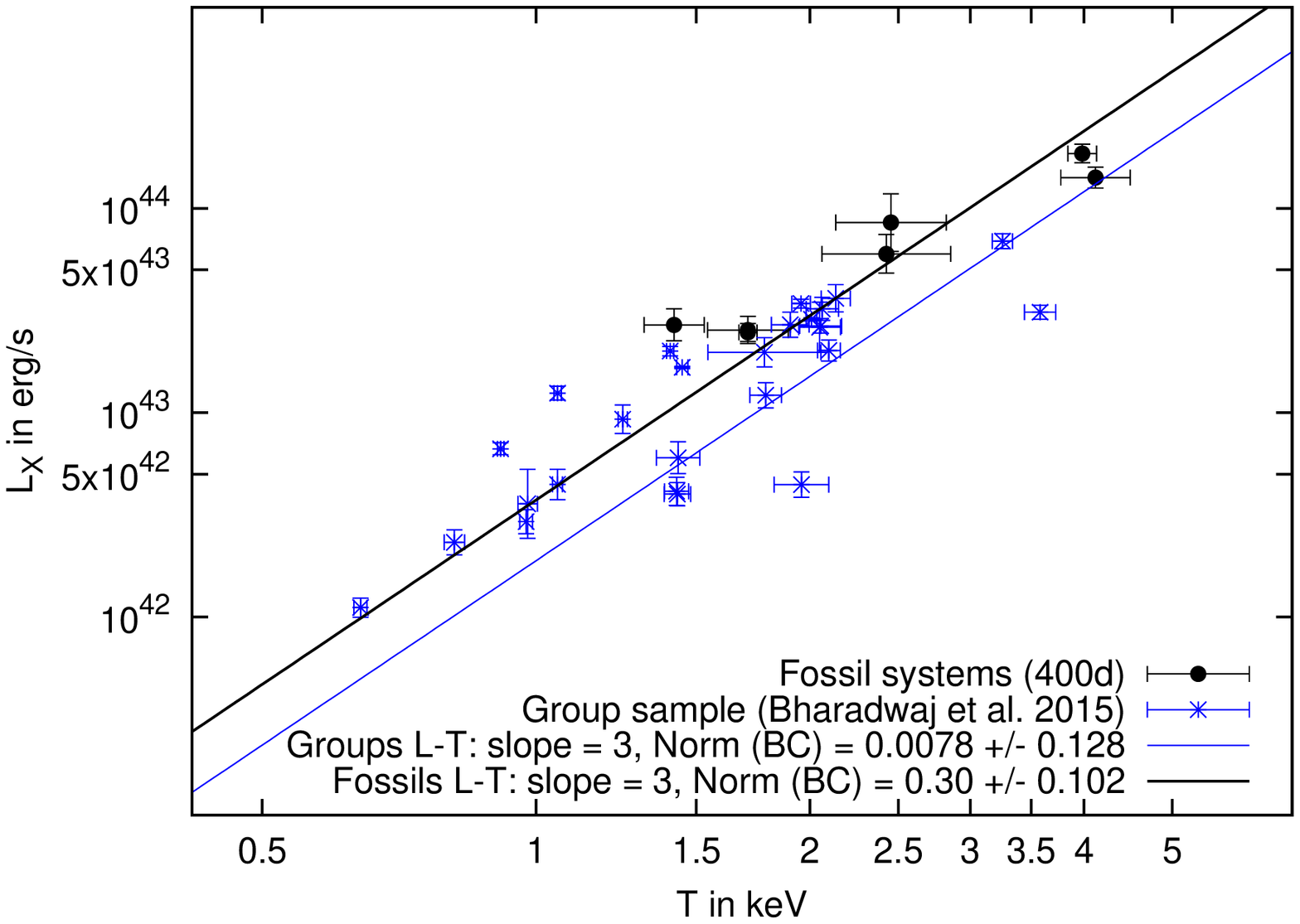}
\caption{\textit{Left:} Observed $L_{\mathrm{X}}-T$ relation for 400d fossils and groups sample. \textit{Right:} Comparison of 400d fossil $L_{\mathrm{X}}-T$ relation and groups relation with frozen slope and bias-corrected normalisations.}\label{LTfossils}
\end{figure*}

This gives a slope of $2.09\pm0.24$ and normalisation (log) of $0.29\pm0.04$. The observed scaling relation for groups in the \cite{2015A&A...573A..75B} has a slope of $2.17\pm0.26$ and normalisation of $-0.01\pm0.09$. To ensure that our choice of region for the temperature determination was not biasing our results, we performed a test by estimating the temperatures in a fixed region of 3 arcmin for all the 400d fossils (effectively corresponding to a median of $\sim 0.5r_{500}$) and redid the scaling relation. No significant changes were found. The observed scaling relation gives us first indications that fossils seem to be more X-ray luminous for a given T (Fig.~\ref{LTfossils}, left), relative to non-fossils, though this could simply be due to selection effects. 

To investigate this further, we performed an additional test where we froze the slopes for both the fossils and groups to 3.0, and only fit the normalisation for both samples. The value of 3.0 for the slope was chosen as generally most groups and clusters can be described well by this particular slope, after factoring in selection effects. The fossils now have a normalisation of $0.38\pm0.11$ and the groups sample have a normalisation of $0.22\pm0.13$. We now proceeded to remove the influence of selection effects on the normalisation of the scaling relation for both samples by generating mock samples of objects like in \cite{2015A&A...573A..75B} by varying the input normalisations. For each mock sample, the slope was always fixed to 3, and the intrinsic scatter fixed to the observed values. Flux and luminosity cuts were applied to both the 400d sample and the groups sample to match the true sample of objects. For the 400d fossil sample, the selection criteria were taken from \cite{2010ApJ...708.1376V}, and can be described as follows: a lower flux cut of $1.4\times~10^{-13}~\mathrm{erg/s/cm}^{2}$, an upper redshift cut of 0.2, and a lower luminosity cut of $10^{43}$ erg/s. Here, fluxes and luminosities are in the ROSAT (0.5--2.0 keV) band. After performing the bias-corrections, the fossils scaling relation has a normalisation of $0.30\pm0.10$ vs. $0.0078\pm0.13$ for the groups, indicating that the large normalisations seem to persist even after accounting for selection effects. Despite a nearly $2.3\sigma$ higher normalisation for fossils with respect to non-fossils, we still treat this finding only as an indication, because on top of the statistical uncertainties on both quantities (which is quite large currently for the 400d fossils), there could also be an additional systematic uncertainty introduced due to differences in the luminosity determination in the different parent catalogues. Secondly, the archival nature of this sub-sample of 400d objects is biased towards systems lacking a SCC (table~\ref{CCprop}) and assuming that the remaining systems are SCC, then adding these objects could potentially increase the normalisation for the scaling relation, and the difference between the fossils and the groups sample would be higher than what we demonstrate here. We plan to explore this in greater detail in a future study of fossils scaling relations.

\begin{table*}
\centering
\caption{Comparison of the ``bias-corrected'' scaling relations for the 400d fossils and the \cite{2015A&A...573A..75B} sample by freezing the slope to 3.}\label{LTtab}
\begin{tabular}{|c|c|c|c|}
\hline \hline
Sample & Slope & Normalisation (observed) & Normalisation (bias-corrected) \\ \hline
400d fossils & $3$ &$0.38\pm0.10$&$0.30\pm0.10$ \\
Groups sample & $3$ &$0.22\pm0.13$ &$0.0078\pm0.13$ \\ \hline
\end{tabular}

\end{table*}

\subsection{Discussion}\label{Disc}
Though our results concern mostly the cores of fossil systems, these properties offer us interesting insights into their formation and evolution. Fossil systems not having experienced any recent major merger activity should have a pronounced cool-core with very short CCTs and `typical' temperature profiles with a clear central temperature drop. The distribution of the CCTs and the shapes of the observed temperature profiles does not however seem to support this idea. Given that the BCG-EP separation is not large and that most fossils still have an EP-EWC separation less than 50 kpc, it seems unlikely that invoking mergers can easily explain the thermodynamic properties of the fossils. 

A lot of these properties can be explained due to a non-gravitational process, in particular the AGN activity at the centres, which we speculate was much more powerful at an earlier epoch than it is now. Simulation results by e.g.~\cite{2007MNRAS.382..433D} and \cite{2011A&A...527A.129D} point out that fossils assemble most of their masses at high redshifts. Interestingly, \cite{2008A&A...490..965D} point out that while the virial mass of the groups were assembled at high redshifts, much of the BCG mass is assembled at much later epochs. Recent results by \cite{2012A&A...537A..25M} conclude that the BCGs of fossils underwent major dissipational mergers at earlier epochs, albeit most of the stellar content was assembled via dissipationless major mergers at later times. Thus, we speculate that during the relatively `gas-rich' phase of mass assembly, the SMBH was strongly fuelled which led to strong AGN feedback such that an almost isothermal central region, or a centrally rising temperature profile was formed. When this phase ended, dissipationless major mergers continued to build the BCG, but there wasn't enough gas to fuel the SMBH strongly, pushing it into a state of relative quiescence. Eventually, for some objects, a small cool-core region was formed probably from the mixture of stellar mass ejecta and the surrounding ICM, not quite unlike thermal coronae seen in groups/clusters. This cool gas started feeding the SMBH, and is now probably involved in a self-regulated feedback cycle which leads to a much lower AGN activity. Four out of the five $z < 0.05$ fossils show extremely weak radio sources \citep{2014A&A...572A..46B,2014MNRAS.444..651M,1998AJ....115.1693C} indicating that the AGN activity is definitely not strong currently, giving credence to this theory. Interestingly, simulation results by \cite{2008ApJ...675.1125B} indicate that non-cool core clusters (cool cores here are defined by a central temperature drop in the temperature profile) at low redshifts ($z < 0.3$) have less halos in their neighbourhood as compared to CC clusters, while the situation was exactly reversed at $z > 1$. Though \cite{2008ApJ...675.1125B} only discuss about clusters with virial masses greater than $10^{14}~\mathrm{M}_{\odot}$, the under dense regions of fossil systems relative to other systems would seem to qualitatively agree with the idea that the progenitors of fossils originated in a region with a higher density of halos that strongly contributed to mergers, and the final mass assembly currently resides in a relatively under-dense region. This is also qualitatively in agreement with the observational results of \cite{2012ApJ...752...12H} who base their conclusions via a study of the fossil ellipticals and the simulation results of \cite{2011A&A...527A.129D}.

The above speculation however, still does not explain the `classical' temperature profile seen for cl1159p5531 and the high temperature fossils. One possibility is that the thermodynamic properties of the fossil systems are strongly dependent on the environment in which it was formed and it could be that these anomalous objects grew in a region relatively under-dense in halos that led to the formation of a proper cool-core via hierarchical, small-scale mergers \citep{2008ApJ...675.1125B} and led to relatively gentle AGN feedback throughout. Secondly, we don't rule out the possibility of the high temperature objects being falsely identified as fossils as the \cite{2009AN....330..978E} sample is largely based on the SDSS catalogue and it has been known for some systems to be wrongly classified as fossils in similar studies, only to be corrected later when better data was available (e.g.~\citealt{2014A&A...565A.116Z} vs.~\citealt{2007AJ....134.1551S}). Cl1159p5531 has a galaxy just outside the search radius which if included would not satisfy the magnitude criterion of \cite{2010ApJ...708.1376V} and be classified as a fossil in that study. Naively though, it seems unlikely that subtle changes in the magnitude gap would have a drastic impact on the ICM properties of these objects.

\section{Summary}\label{Summary}
We have studied the core properties of 17 fossil systems from literature with data from the Chandra archives. The conclusions of our work can be summarised as follows:
\begin{itemize}
\item Most fossil systems (82\%) are identified as cool-core objects based on at least two diagnostics. Interestingly though, there are indications that most fossil systems are WCC objects and that the population of non-SCC fossils outstrip SCC fossils.
\item Fossil systems show a large range of temperature profiles with many cool-core objects lacking the expected clear central temperature drop.
\item We analysed the X-ray emission coincident with the BCG for fossils with $z<0.1$ and conclude that these are not the typical X-ray coronae reported for some other objects. We speculate that the stellar X-ray component (if present) has mixed with the ICM.
\item We performed a deprojection analysis using the SBPs for fossils with $z < 0.05$ and derived their thermodynamic properties. These objects lack group-sized cool-cores and show evidence for non-gravitational processes (AGN feedback in particular) based on the shallow powerlaw indices of their entropy profiles.
\item There are some indications that the normalisation of the $L_{\mathrm{X}}-T$ relation is higher for fossils than for non-fossils.
\item We speculate that early, dissipational major mergers led to a strong fuelling of the central SMBH leading to powerful AGN feedback and this could explain the lack of group-sized cool-cores in low-temperature fossil groups. There are however anomalous cases which lead us to conclude that this is probably an incomplete picture of the formation and evolution of fossil systems.
\end{itemize}

In short, we have offered a glimpse into the nature of fossil systems through their core properties. In order to concretise these ideas, we will have to construct large, objectively selected fossils sample with high quality X-ray data; ideally with some mass/luminosity cut to distinguish between fossil groups and clusters. A key point here seems to be to also improve our understanding of the weak cool-core cluster/group regime, which till date has been probed poorly. These observational results will also need backing from detailed, high-resolution simulations which factors in feedback processes that would help us unravel substantial features in the growth of fossil systems and the impact on the properties of the ICM.

\begin{acknowledgements}
{The authors would like to thank the referee for constructive feedback. V.~B. would like to thank Lorenzo Lovisari for constructive comments. V.~B. and G.~S. acknowledge support from the DFG via grant RE 1462/6. T.~H.~R. acknowledges support from the DFG through the Heisenberg research grant RE 1462/5. The program for calculating the CCT was kindly provided by Paul Nulsen, which is based on spline interpolation on a table of values for the APEC model assuming an optical thin plasma by R. K. Smith. This work has made use of the NASA/IPAC extragalactic database (NED) which is operated by the Jet Propulsion Laboratory, California Institute of Technology, under contract with the National Aeronautics and Space Administration.}
\end{acknowledgements}

\bibliographystyle{aa}
\bibliography{ref}

\appendix
\onecolumn
\section{Temperature Profiles}\label{Tempprofiles}
\begin{figure*}[h!]
\includegraphics[angle=270,scale=0.50]{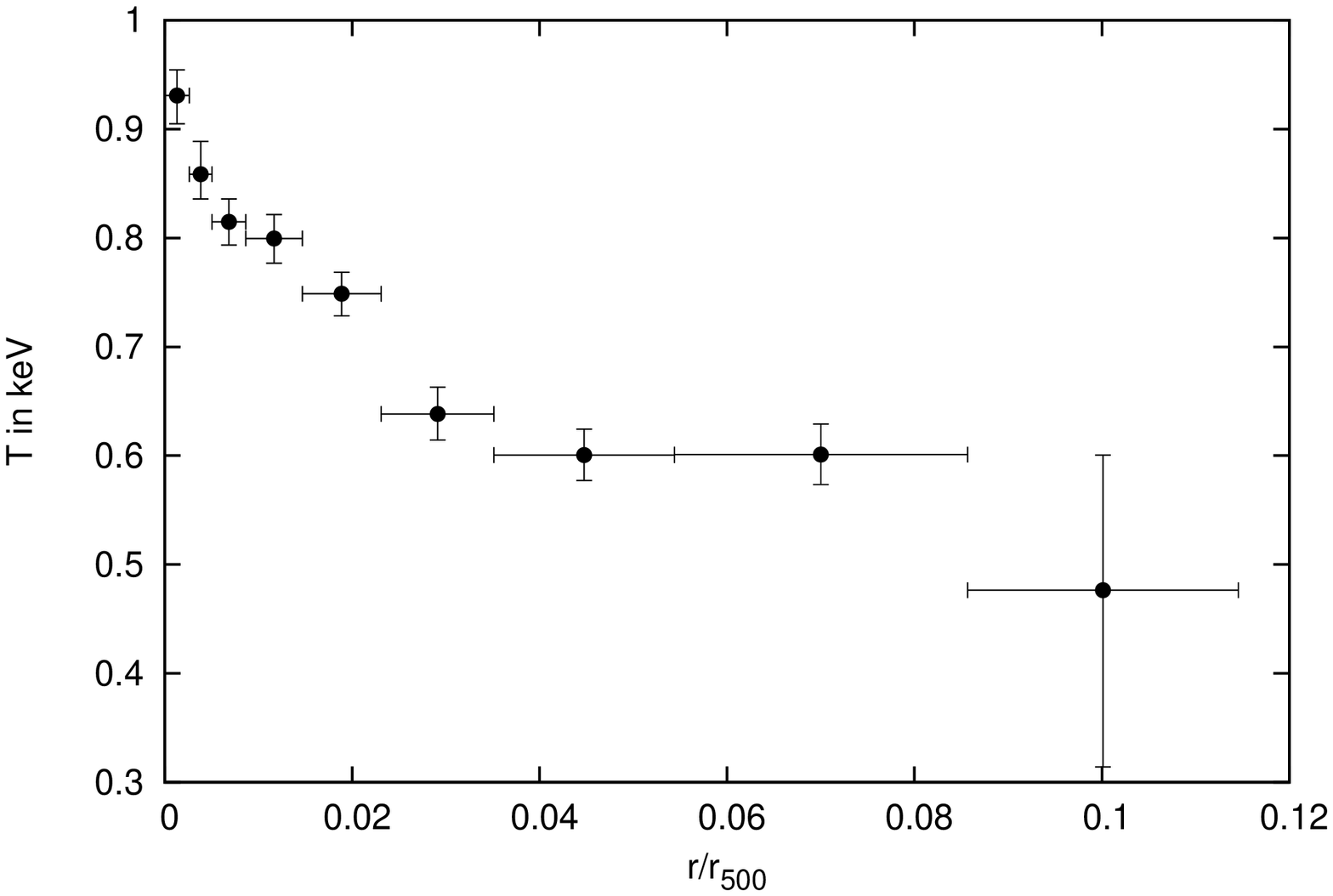}
\includegraphics[angle=270,scale=0.50]{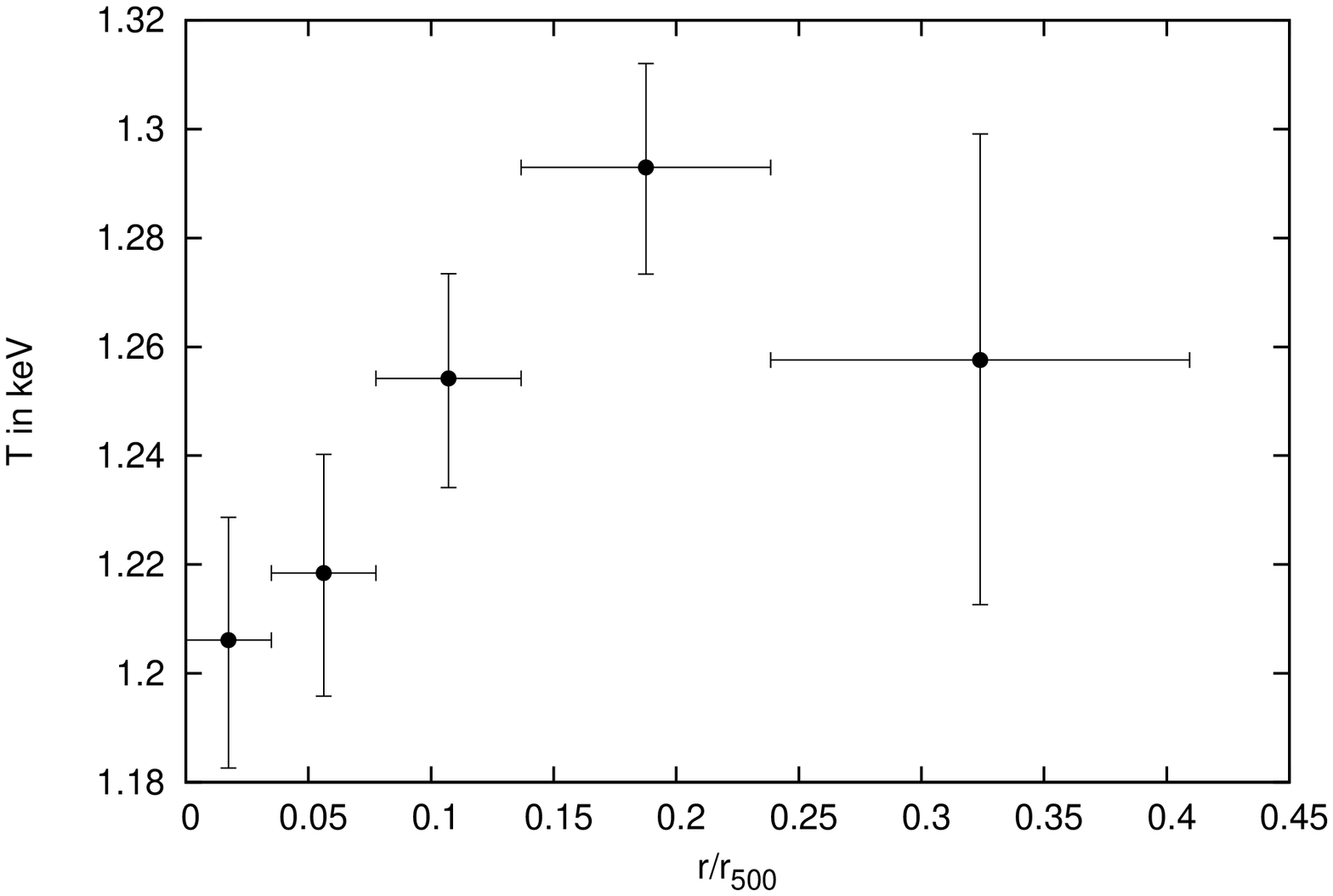}
\caption{Temperature profiles of NGC 6482 and NGC 1132.}
\end{figure*}

\begin{figure*}[h!]
  \includegraphics[angle=270,scale=0.50]{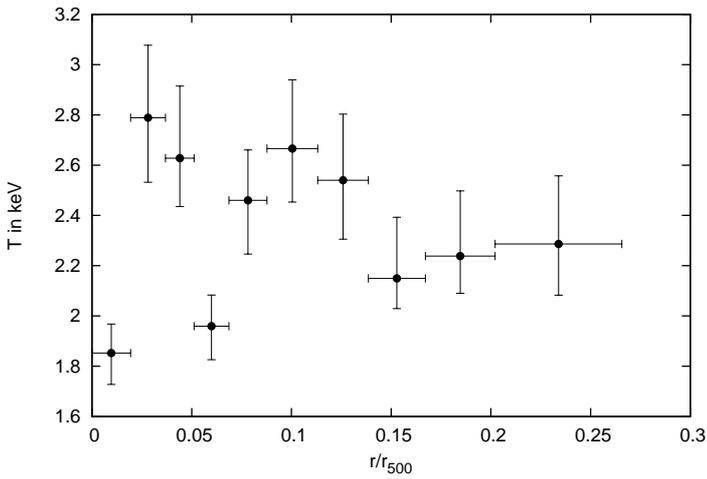}
\includegraphics[angle=270,scale=0.50]{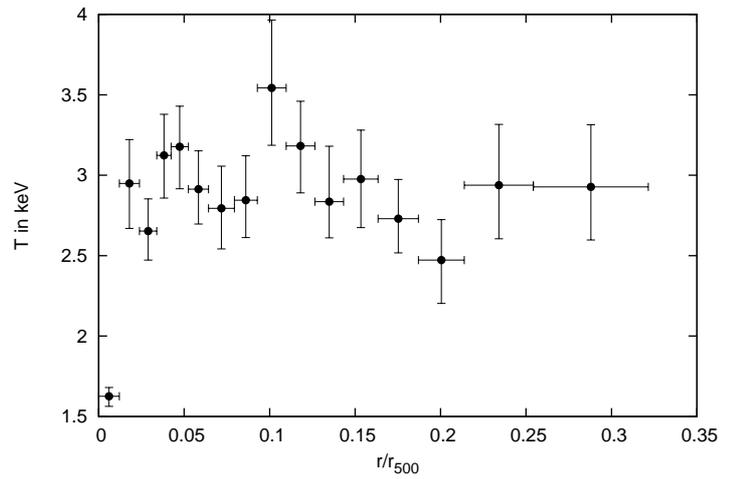}
\caption{Temperature profiles of RX J0454.8-1806 and ESO306.}
\end{figure*}

\begin{figure*}[h!]
\includegraphics[angle=270,scale=0.50]{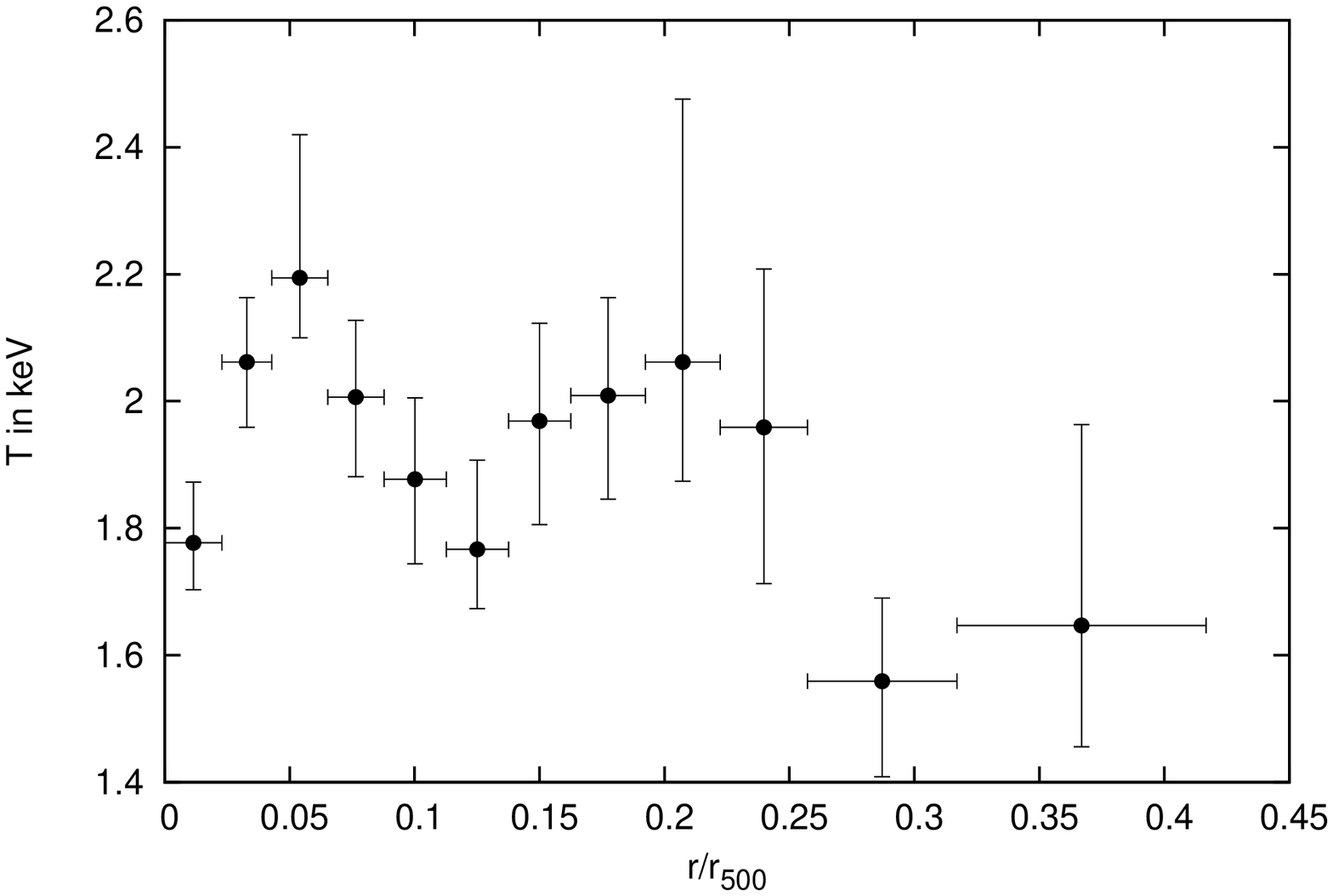}
\includegraphics[angle=270,scale=0.50]{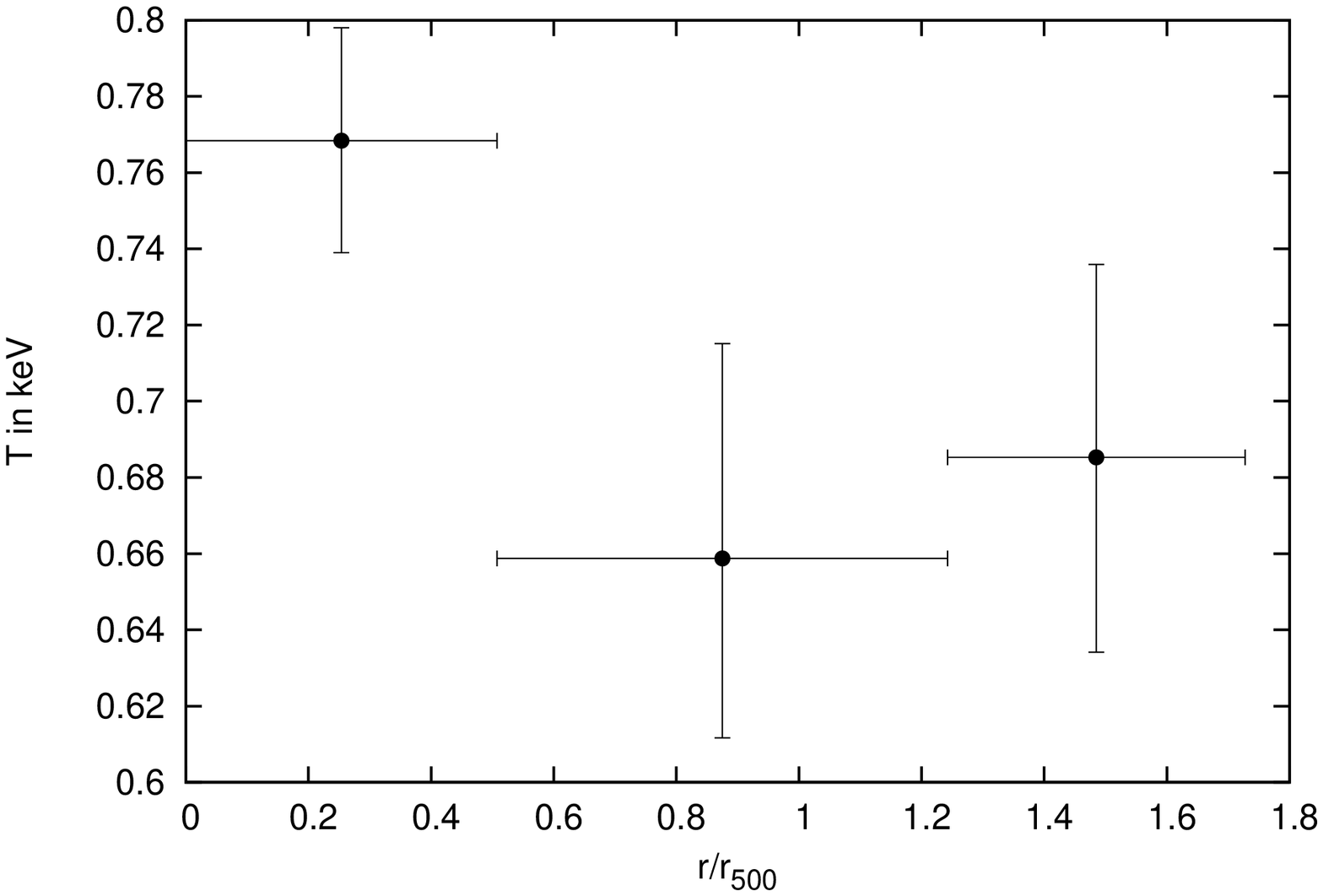}
\caption{Temperature profiles of UGC 842 and RX J1331+1108. For RX J1331+1108, the blank-sky files might be under-estimating the background beyond $r_{500}$. This should however not be a problem in the central regions.}
\end{figure*}

\begin{figure}[h!]
\includegraphics[angle=270,scale=0.50]{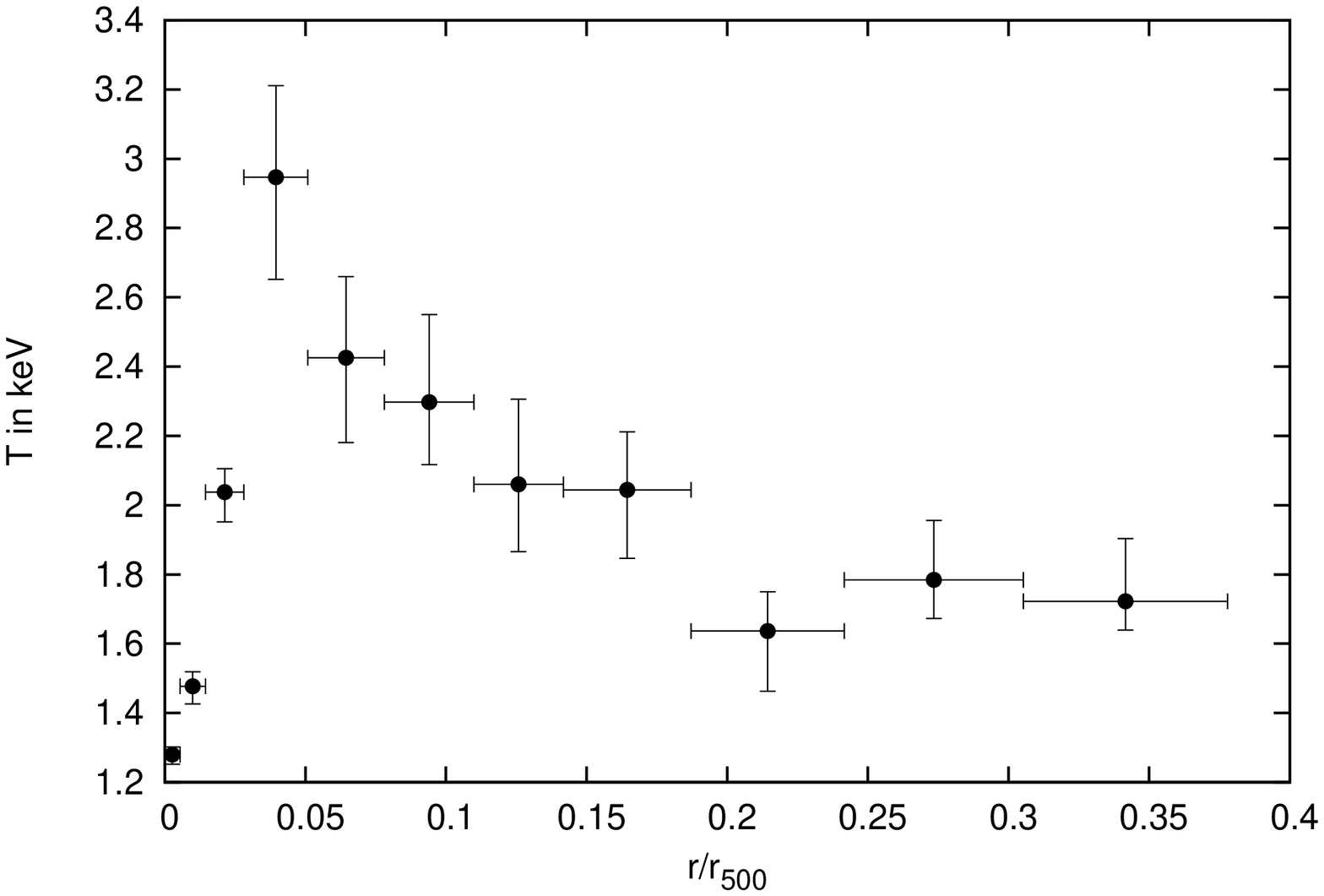}
  \includegraphics[angle=270,scale=0.50]{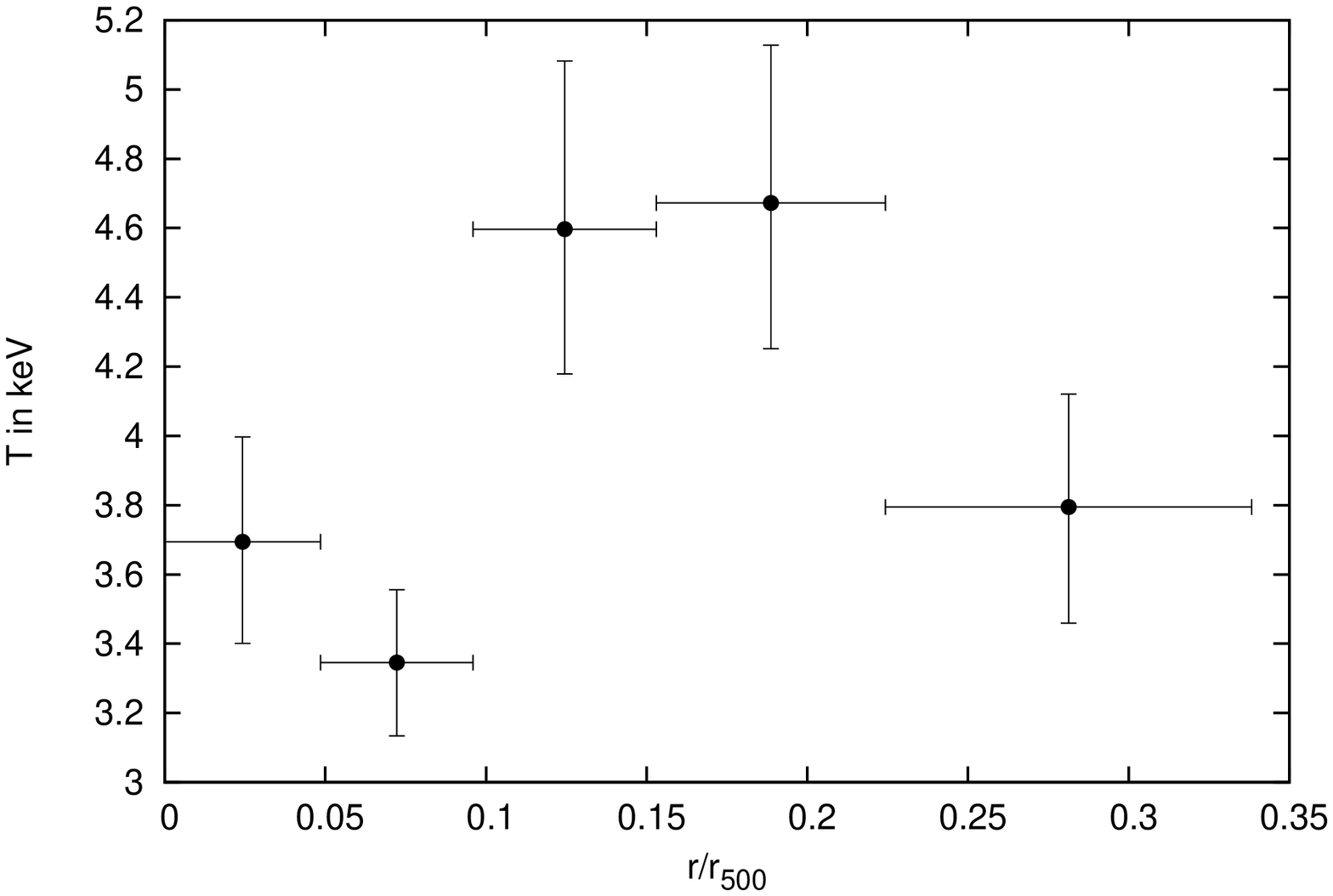}
\caption{Temperature profiles of cl1159p5531 and cl1416p2315.}
\end{figure}

\begin{figure}[h!]
\includegraphics[angle=270,scale=0.50]{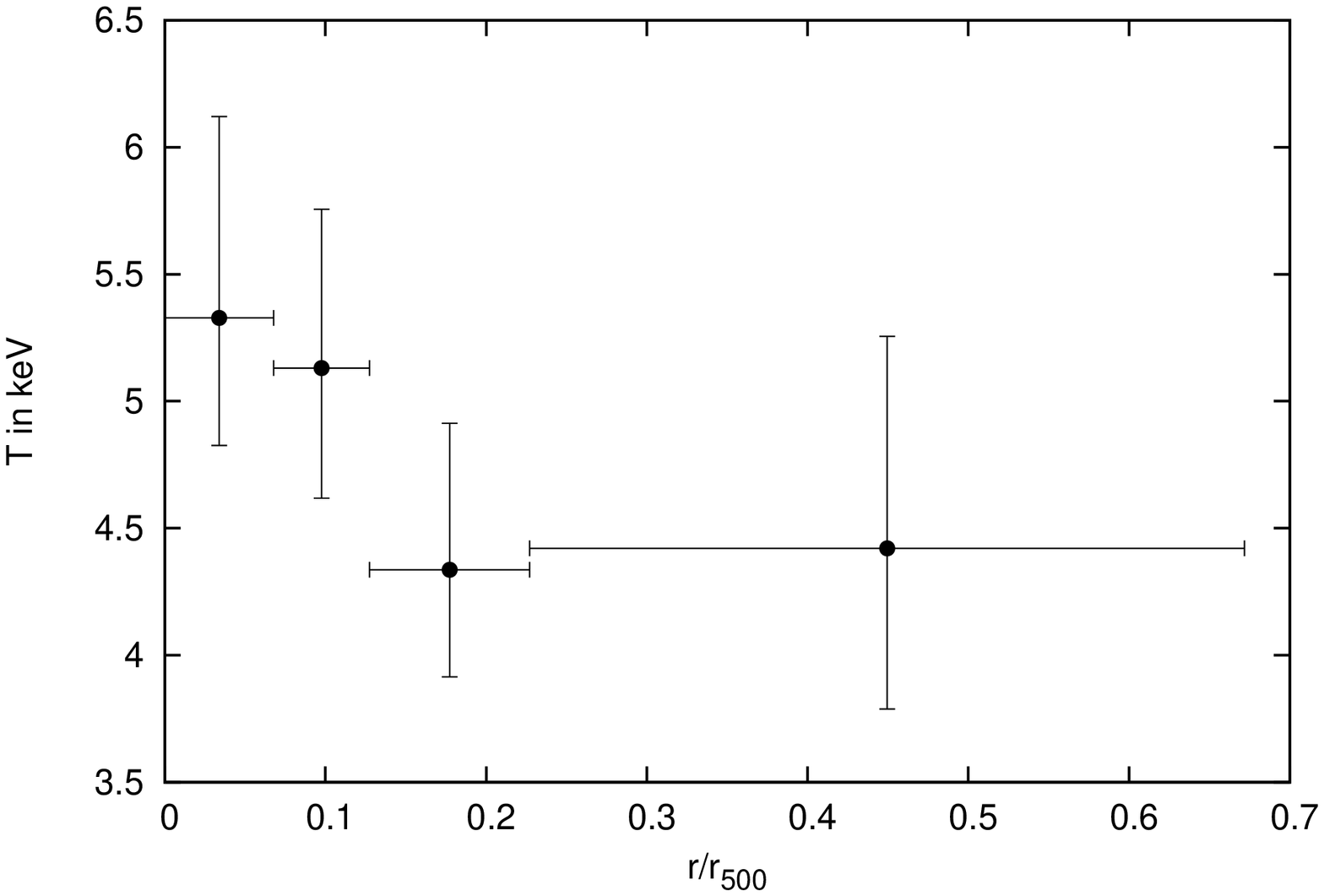}
\includegraphics[angle=270,scale=0.50]{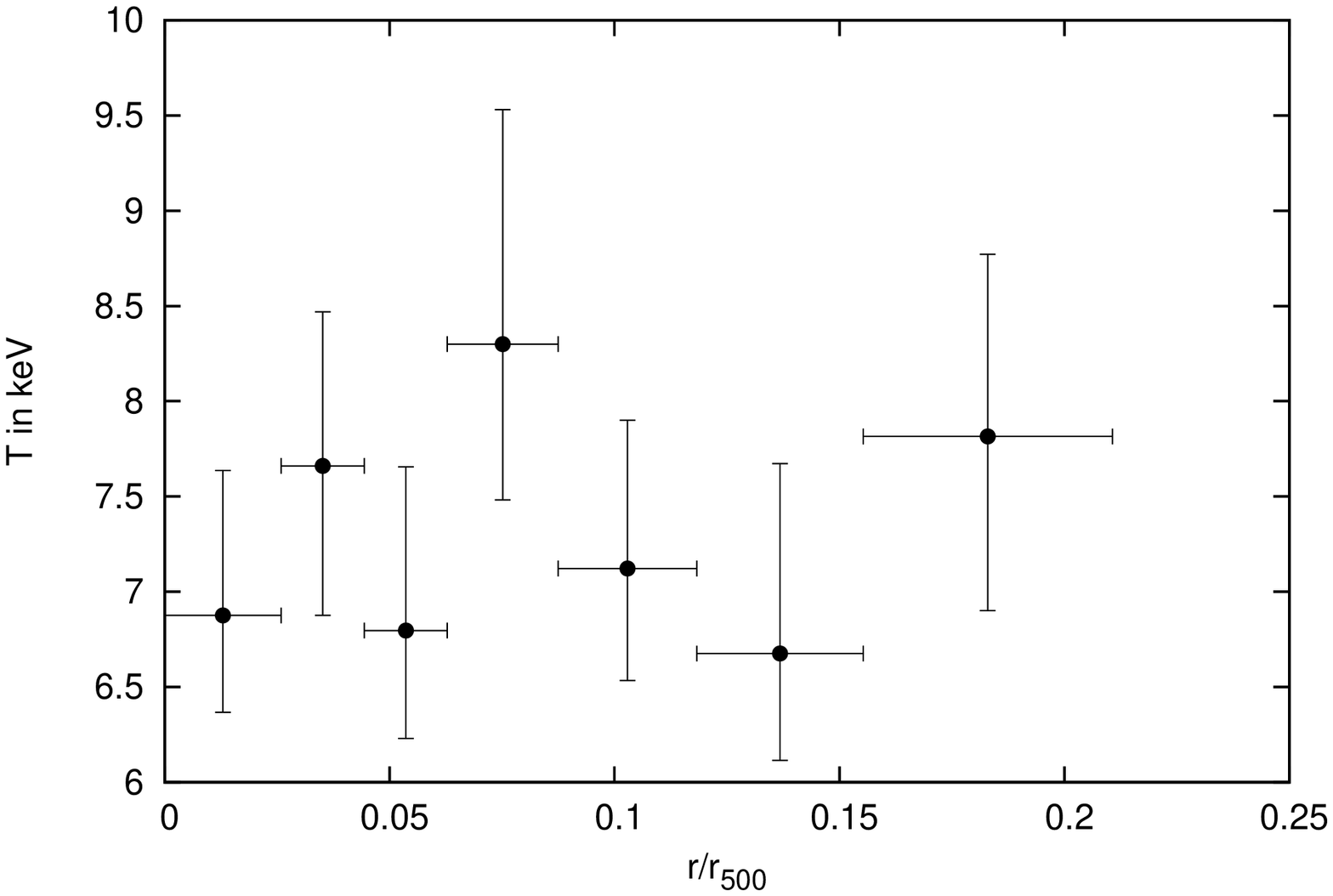}
\caption{Temperature profiles of RX J0825.9+0415 and RX J0801+3603.}
\end{figure}

\begin{figure}[h!]
\includegraphics[angle=270,scale=0.50]{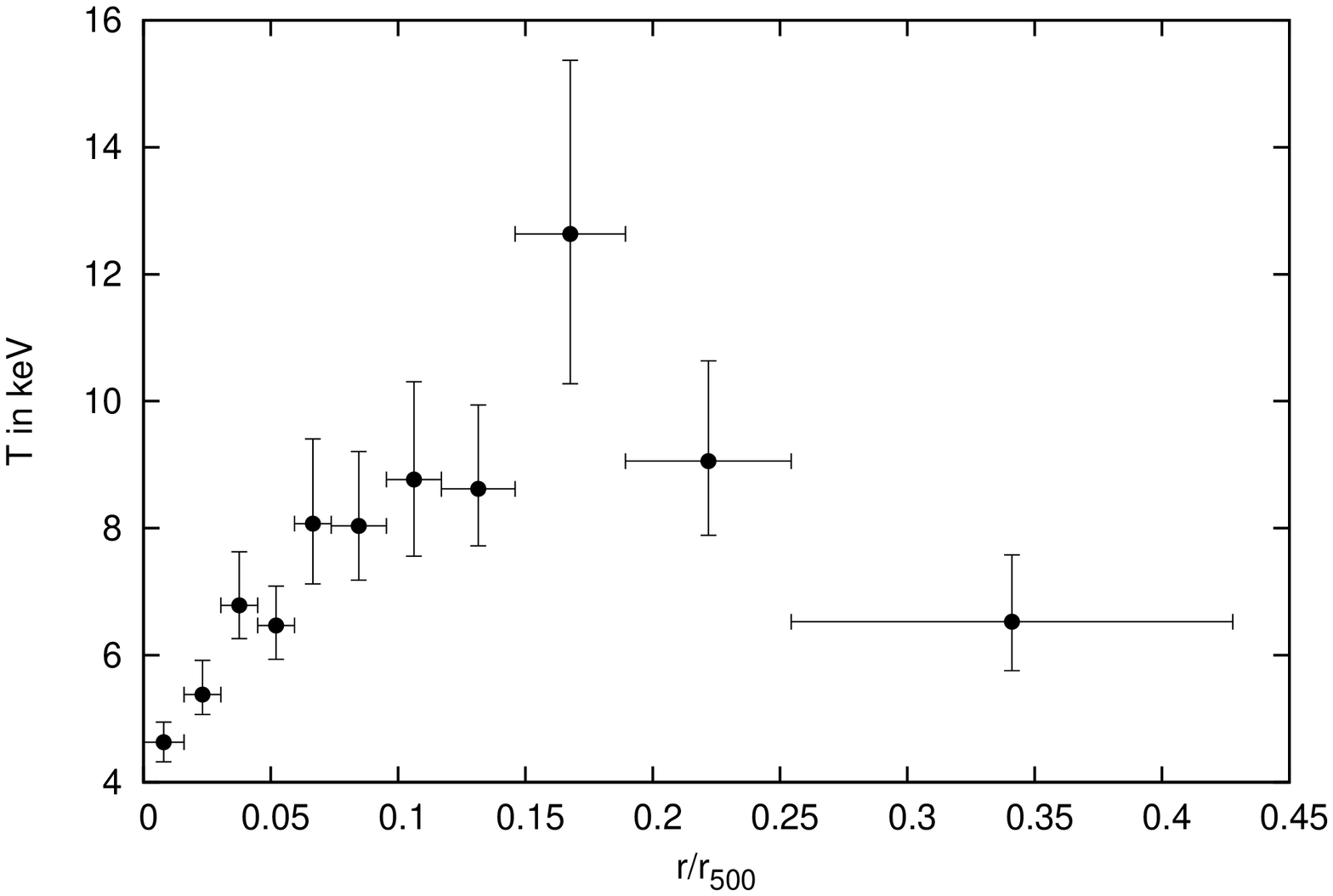}
 \includegraphics[angle=270,scale=0.50]{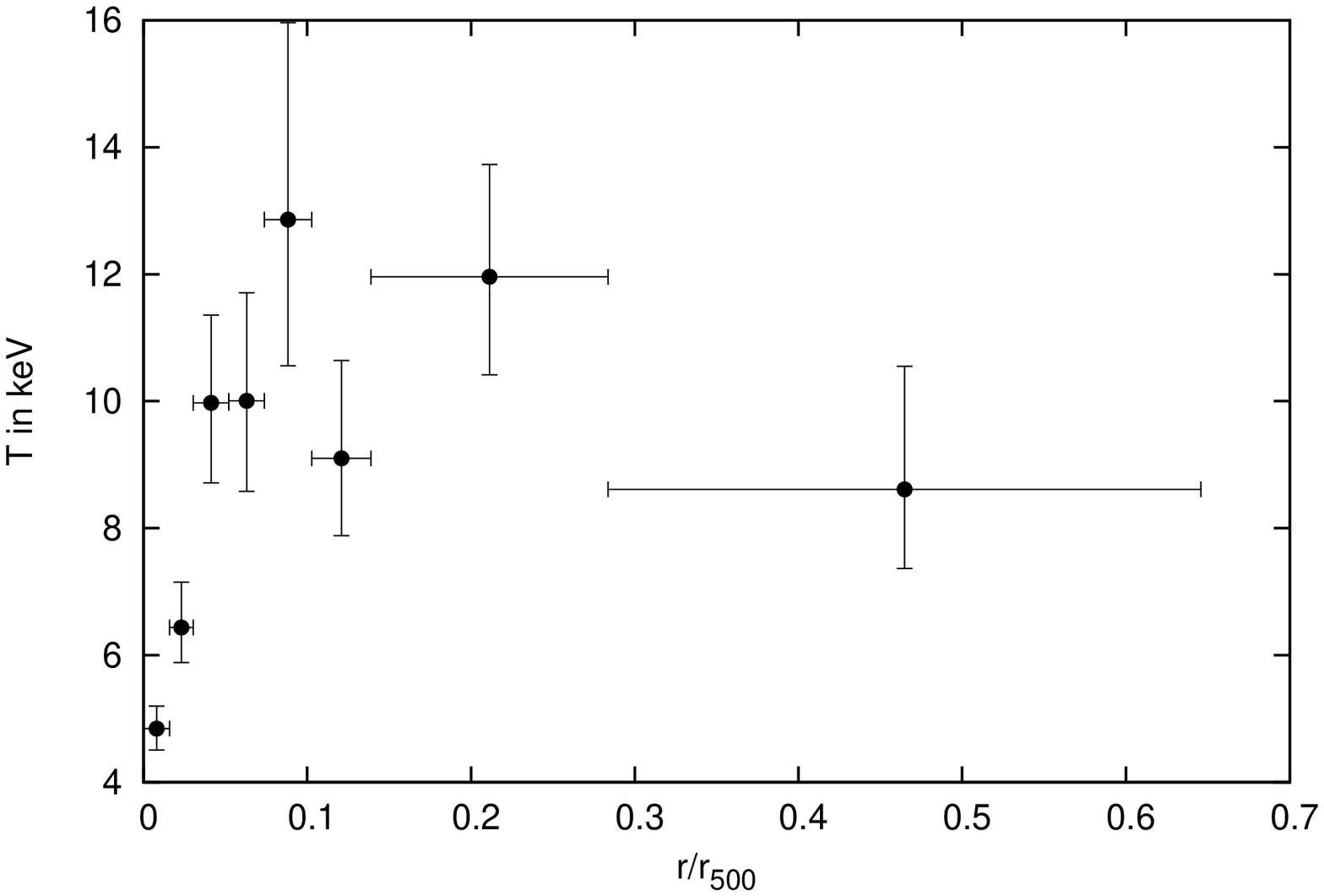}
\caption{Temperature profiles of RX J1115.9+0130 and RX J0159.8-0850.}
\end{figure}

\end{document}